\documentclass[sigconf]{acmart}
\usepackage{graphicx}
\usepackage{CJKutf8}
\usepackage{enumitem}
\usepackage{xcolor} 
\usepackage{xspace}
\usepackage{graphicx}

\newcommand{\SysName}{\textsc{SherAgent}\xspace}
\usepackage{subcaption} 
\usepackage{multirow}
\usepackage{tabularx}
\usepackage{booktabs}
\usepackage{makecell}
\usepackage{array}
\usepackage{tabularx}
\usepackage{booktabs}
\usepackage{graphicx}
\usepackage{placeins}
\usepackage{xcolor}

\setcopyright{acmlicensed} 
\copyrightyear{2018} 
\acmYear{2018} 
\acmDOI{XXXXXXX.XXXXXXX} 
\acmConference[Conference acronym 'XX]{Make sure to enter the correct
  conference title from your rights confirmation email}{June 03--05,
  2018}{Woodstock, NY}  
\acmISBN{978-1-4503-XXXX-X/2018/06}  

\begin{document}
\author{Zhenyuan Li}
\affiliation{
  \institution{Zhejiang University}
  \country{China}
}

\author{Zhengkai Wang}
\affiliation{
  \institution{Zhejiang University}
  \country{China}
}

\author{Ling Jiang}
\affiliation{
  \institution{Tencent Security Keen Lab}
  \country{China}
}

\author{Xiangmin Shen}
\affiliation{
  \institution{Hofstra University}
  \country{USA}
}

\author{Ruixiao Lin}
\affiliation{
  \institution{Zhejiang University}
  \country{China}
}

\author{Sen Nie}
\affiliation{
  \institution{Tencent Security Keen Lab}
  \country{China}
}

\author{Shi Wu}
\affiliation{
  \institution{Tencent Security Keen Lab}
  \country{China}
}

\author{Shouling Ji}
\affiliation{
  \institution{Zhejiang University}
  \country{China}
}

\title{\SysName: Scaling Attack Investigation in the Wild via LLM-Empowered Iterative Query-Filter Backtracking}
\begin{abstract}
Provenance-based attack investigation enables viable automation by standardizing data and query logic; however, it is critically hindered in practice by dependency explosions and fragmented causal chains in the wild.
Towards designing a robust and automated investigation tool, we collaborated with the SOC of a major Internet corporation serving billions of users. By engaging in real-world incident response, we are able to evaluate and refine their existing LLM-based investigation workflows, which processes tens of thousands of raw alerts daily, leaving thousands for manual triage, to find out the root causes of investigation failures and major challenges in their existing tools.

Motivated by these findings, we propose \SysName, an LLM-empowered automated investigation system. Operating on an iterative ``query-filter'' backtracking paradigm over provenance graphs, \SysName leverages the semantic reasoning capabilities of LLMs to process unstructured data, such as investigation context and threat intelligence. To overcome fragmented causal chains caused by missing events, the system dynamically calibrates query conditions to broaden the search scope. Concurrently, it performs precision result filtering and strategic nodes selection for subsequent exploration, thereby mitigating dependency explosions. 
Extensive evaluations in the wild demonstrate that \SysName improves the end-to-end investigation success rate by 31.1\% and 63.7\% compared to both legacy enterprise baselines and SOTA approaches, respectively. Furthermore, it operates with remarkable efficiency, incurring under \$0.10 in API costs and requiring less than 4 minutes per investigation. Finally, our user study confirms that \SysName provides accurate and clear insights, significantly reducing the analytical overhead for security experts.
\end{abstract}


\maketitle

\section{Introduction}
\label{sec:intro}
Timely attack investigation is critical for containing threats and mitigating subsequent breaches; however, traditional manual triage is prohibitively expensive and unscalable. Large-scale organizations and public cloud vendors face a pervasive challenge in alert fatigue~\cite{10.1145/3243734.3243794, 192408}, primarily driven by the parallel deployment of disparate intrusion detection systems. These systems generate a massive volume of security alerts characterized by inconsistent quality and fragmented contextual information~\cite{9152771, 10.1145/3723158}, rendering the operational overhead of manual investigation nearly insurmountable.
Despite these challenges, inadequately investigated alerts introduce severe security risks~\cite{8835390, BUSETTI2025102000}.
An illustrative example is the 2020 Toll Group ransomware incidents: the company’s failure to identify the initial point of compromise in early 2020 allowed attackers to persist, leading to a second, more devastating ransomware attack (Nefilim) in May that exfiltrated 200 GB of data\footnote{Exfiltrate, encrypt, extort, https://www.aspi.org.au/report/exfiltrate-encrypt-extort/}. This case underscores that failure to resolve the root cause through rigorous investigation inevitably results in repeated compromises.

Recently, provenance analysis has emerged as a promising solution, transforming complex manual investigations into standardized paradigms of iterative querying and filtering~\cite{LI2021102282, 10.1145/3539605, 190900, 9900181}. As Table~\ref{tab:systemComparison} shows, existing works generally follow two approaches: backward-tracking-based reconstruction, which prunes graphs to highlight attack paths, and query-based hunting, which relies on predefined templates to identify threats. Backward-tracking-based methods, in particular, assist reconstruction by reducing graph scale through impact weighting or anomaly detection~\cite{263852, 9833671, 10.1145/945445.945467, king2005enriching, Kwon2018MCIM, Lee2013HighAA, 10.1145/3133956.3134045,Liu2018TowardsAT, 277080,  Hassan2019NoDozeCT, 10.1145/3133956.3134015, wei2026rapidle, 291066, 10.1145/3719027.3765219}.
However, these methods face two critical challenges in production environments. First, data incompleteness is inevitable; various factors, including lossy data compression, unreliable log collection, and attacker anti-forensic operations, can all result in missing nodes or edges. Such gaps fragment causal chains and render full reconstruction impossible. Second, traditional methods rely primarily on low-level system behaviors and lack high-level semantic reasoning. This \textit{semantic gap} makes it difficult to distinguish malicious activities from benign system functions that exhibit similar dependencies, leading to investigation reports polluted by false positives. Especially when traversing highly connected ``hub'' nodes, the absence of semantic guidance triggers the notorious \textit{dependency explosion} problem~\cite{9152772}.

Query-based methods were originally conceived to expedite investigations by optimizing data retrieval throughput. While these approaches can achieve high precision through manual expert-driven query formulation, they suffer from prohibitive operational overhead~\cite{217496, 215975}. Although automated query schemes have emerged to alleviate this labor burden, they rely primarily on rigid, pre-defined templates and historical heuristics, which severely limit their adaptability to novel or polymorphic threats~\cite{9458828}.
Recently, Security Operations Centers (SOCs) have begun adopting LLM-based investigations to gain deeper semantic insights into system activities~\cite{aisoc}. However, general-purpose models often lack the specialized domain expertise required for technical forensics and are susceptible to contextual hallucinations during long-term, multi-stage analysis. Furthermore, the inherent non-determinism of LLMs introduces significant inconsistency, hindering the reproducibility and reliability essential for automated forensics.

To motivate and facilitate our subsequent design, we collaborated with the SOC of a major Internet corporation serving billions of users. By engaging in real-world incident response, we evaluated and refined their existing LLM-based investigation workflows by conducting a semi-manual empirical study of over 10,000 security alerts collected from diverse operational environments, including internal corporate networks, cloud infrastructure, and integrated security services. 
Our analysis investigated alert typologies, attack characteristics, and the underlying failure modes and bottlenecks of existing naive LLM-based automated investigation methods.
Our findings reveal that the majority of attacks triggering these alerts are structurally simple. As illustrated in Figure~\ref{fig:Overview_of_investigation} (a)(b)(c), attack chains rarely exceed five hops. However, approximately 40\% of these alerts exhibit high recurrence rates (Figure~\ref{fig:Overview_of_investigation} (d)), strongly suggesting that existing investigation systems fail to find the initial access vectors. In severe cases, specific attacks persistently re-trigger alerts over consecutive periods exceeding seven days.
Furthermore, Figure~\ref{fig:type_fail_combined} categorizes the root causes and primary difficulties leading to investigation failures. That is, the predominant challenges bottlenecking current LLM-based systems stem from underlying telemetry data loss and the complex reasoning required to parse unstructured security data. 

In summary, while traditional provenance-based methods cannot handle fragmented audit logs, unconstrained LLMs suffer from semantic drift and dependency explosion due to overly broad querying. The fundamental challenge lies in reconciling causal rigor with adaptive flexibility to maintain investigative continuity across incomplete forensic data.
Motivated by these findings, we propose \SysName, an LLM-based automated attack investigation system built upon iterative SQL querying for relevant data retrieval and semantic log analysis for filtering and investigation branch selection. Our key insight is that logical continuity, embedded within semantic fields (e.g., CmdLines and file paths), can effectively compensate for physical fragmentation in raw logs. Thus, by shifting from traditional graph traversal to the SQL-based discovery of threat-relevant entities and events, \SysName maintains investigation integrity even in the absence of strict causal connectivity.
Specifically, to ensure reliability and efficiency, we introduce three core mechanisms:
1) Template-Based Querying: We constrain the LLM's outputs to predefined SQL templates, effectively transforming unpredictable open-ended generation into causality-guided query formulation, thereby constraining the search space while maintaining operational flexibility.
2) Semantic Pruning: By enriching log context, embedding domain-specific threat knowledge, and systematically tracking investigation progress, the LLM can accurately filter irrelevant branches to suppress dependency explosion.
3) Tree-Structured State Maintenance: We maintain the global investigation state in a dynamic tree structure rather than a rigid Depth-First Search path. This allows the system to continuously re-evaluate evidence and pivot to high-priority branches, preventing the LLM from pursuing erroneous analysis paths and providing the system with essential task rollback capabilities.

\begin{table}[tbp]
\centering
\caption{Comparison with Attack Investigation Systems}
\label{tab:systemComparison}
\footnotesize
\begin{tabular}{cccc}
\toprule
{System} & {\makecell{Unstructured\\data analysis}} & {\makecell{Pre-train\\-free}} & {\makecell{Across\\broken chain}}  \\
\midrule
\textbf{\SysName (Ours)} & \textbf{\checkmark}   &  \textbf{\checkmark}    & \textbf{\checkmark} \\
\makecell{Propagation-based Approaches\\ \cite{277080, king2005enriching, Kwon2018MCIM, Lee2013HighAA, 10.1145/3133956.3134045, Liu2018TowardsAT}}   & $\times$         &  \textbf{\checkmark}    & $\times$            \\
\makecell{Anomaly-based Approaches\\ \cite{9833671, 10.1145/3719027.3765219, 10.1145/945445.945467, Hassan2019NoDozeCT, 10.1145/3133956.3134015, wei2026rapidle, 291066, 263852}}       & $\times$        &  $\times$     & $\times$   \\
Query-based Approaches~\cite{217496,215975, 9458828}         & $\times$     &  $\times$     &  \textbf{\checkmark}      \\
\bottomrule
\end{tabular}
\end{table}

\begin{figure}[tbp]
    \centering
    \includegraphics[width = 0.91\linewidth]{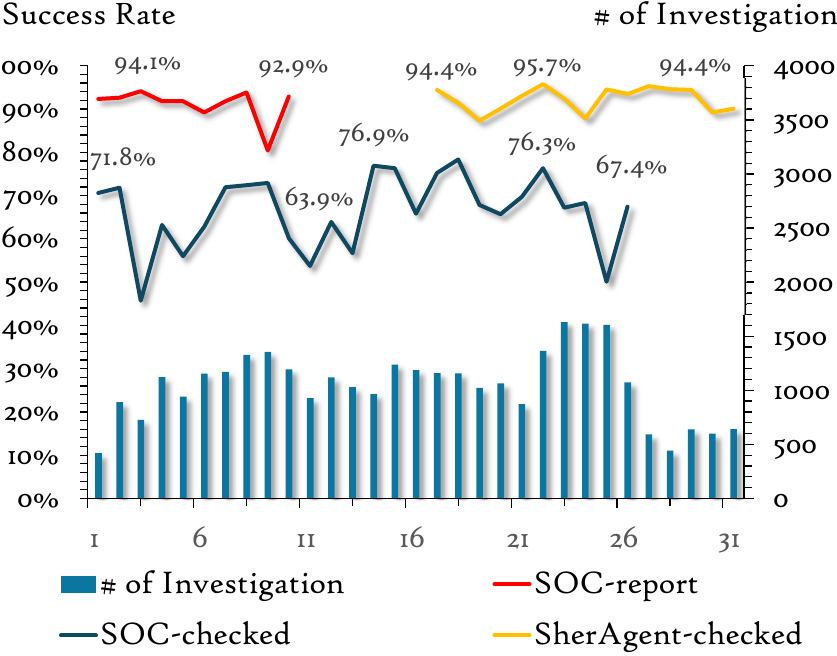}
    \caption{Comparison of End-to-End Investigation Success Rate in the Wild (January, 2026)}
    \label{fig:Application}
\end{figure}

As shown in Figure~\ref{fig:Application}, we conducted a pilot evaluation in January 2026 to benchmark our system against the legacy infrastructure. While the legacy tool reports a deceptively high self-reported success rate (red line), our manual audit reveals its actual verified performance, hereafter referred to as the \textit{SOC-baseline} (blue line), to be only 64.7\%. This discrepancy is primarily driven by contextual hallucinations, where the legacy model generates plausible-sounding but factually incorrect investigation reports that fail human verification. Consequently, its real-world utility falls significantly behind our proposed system.
Following this validation, we deployed \SysName into a live production environment. Over several months of continuous operation, the system has processed 53,849 real-world security alerts, maintaining a verified success rate exceeding 96\% (based on periodic spot-checks, with 10 cases sampled per day, totaling 980 cases, approximately 2\% of all alerts). In the experimental setting, \SysName achieved a success rate of 92.2\%. This performance represents an absolute end-to-end improvement of {{31.1}\% over the \textit{SOC-baseline} and {63.7\%} compared to the academic state-of-the-art (SOTA).
Furthermore, \SysName demonstrates remarkable operational efficiency, incurring an average API cost of less than \$0.10 and requiring under 4 minutes per investigation. A user study involving 10 security practitioners further confirms that \SysName accelerates investigative workflows by approximately 10–20 minutes per case, while earning high qualitative scores for both investigative accuracy and report clarity.

\begin{figure*}[htbp]
  \centering
  \includegraphics[width=0.91\linewidth]{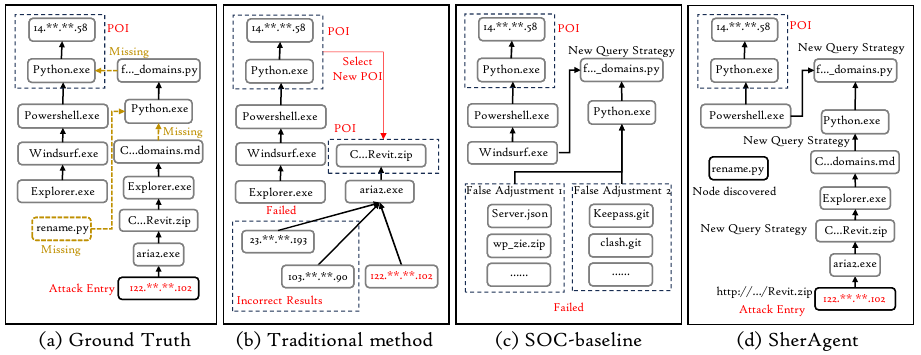}
  \caption{Motivating Examples}
  \label{fig:example}
\end{figure*}

This paper makes the following contributions:

\begin{enumerate}[label=\textbf{C\arabic*.}, left=0pt, itemsep=0em]
\item Through collaboration with a major enterprise SOC, we conducted an extensive empirical study of attack investigations. Our findings identify log data omissions, alongside the notorious dependency explosion and semantic gap, as the primary root causes of investigation failures. Furthermore, we demonstrate that existing investigation approaches inherently struggle with these challenges due to their rigid heuristics and insufficient semantic reasoning capabilities.
\item To address these practical limitations, we propose \SysName, an automated, LLM-empowered provenance investigation system. By leveraging the code generation, semantic analysis, and logical reasoning capabilities of LLMs, \SysName supports a robust ``query-filter'' backtracking paradigm. This architecture allows the system to effectively bridge fragmented attack chains caused by missing events and performs intelligent branch selection, significantly enhancing both investigative accuracy and operational efficiency.
\item We systematically evaluated \SysName through a live production deployment, where it processed 53,849 real-world alerts. The results demonstrate that \SysName effectively reconstructs fragmented causal chains resulting from data omissions and achieves high-precision filtering of irrelevant logs. Compared to industrial baselines, \SysName offers reduced latency, minimal operational overhead, and enhanced interpretability for downstream security operations.
\end{enumerate}

\section{Investigating Attacks in the Wild}
\label{sec:background}

Provenance-based backtracking is a fundamental technique in modern attack investigations. As illustrated in Figure~\ref{fig:example}, the investigation process typically originates from a localized alert, which serves as the Point of Interest (POI). Starting from this POI, the analyst or system queries the stored provenance logs, where system activities are modeled as tuples of $\langle$\textit{source entity}, \textit{sink entity}, \textit{relationship}, \textit{timestamp}$\rangle$. By recursively traversing these logs along reverse causal dependencies, analysts can iteratively reconstruct the sequence of entities and events that constitute the underlying attack chain.

To investigate the practical challenges of threat analysis in real-world environments, we collaborated with the Security Operations Center (SOC) of a leading global network service provider. This department is tasked with securing large-scale cloud infrastructures, internal corporate networks, and managed security services for external clients. Their security ecosystem features a mature data infrastructure capable of collecting and storing fine-grained provenance data in compliance with the OCSF (Open Cybersecurity Schema Framework) standard~\cite{ocsf}. Leveraging this collaboration, we conducted an empirical study on over 10,000 high-severity alerts, distilled from hundreds of thousands of raw security events collected in 3 months, across diverse cloud-based Linux and enterprise Windows environments. Our distillation strategy specifically focused on cases where the legacy SOC system yielded suboptimal investigative outcomes. Furthermore, we included all cases where \SysName and the SOC team conducted joint investigations during the testing phase. Our study systematically evaluates alert taxonomies, investigation success rates, and the underlying root causes of investigation failures.

\subsection{Statistics of Alerts}

\begin{table}[t]
\centering
\footnotesize
\caption{Classification and An Example of System Log}
\label{tab:initial_info}
\begin{tabularx}{\linewidth}{l X X}
\toprule
\textbf{Log} & \textbf{Process $\rightarrow$ X} & \textbf{X $\rightarrow$ Process} \\
\midrule
Process 
& Launch, Inject, Set User ID 
& Terminate, Open \\
Network 
& Reset, Traffic 
& Close, Fail, Refuse, Listen, Open \\
File 
& Create, Update, Delete, Rename, Set Attributes, Set Security, Encrypt, Mount 
& Read, Get Attributes, Get Security, Decrypt, Unmount, Open, Close \\
\bottomrule
\end{tabularx}
\begin{tabularx}{\linewidth}{l X}
\toprule
\textbf{Example} & \textbf{Fields} \\
\midrule
Activity & Network \\
Types & Traffic \\
Source & 
\textbf{Type:} process; 
\textbf{Name:} svchost.exe; 
\textbf{CmdLine:} \texttt{svchost.exe -k NetworkService -p}; 
\ldots \\
Sink &  
\textbf{Type:} endpoint; 
\textbf{Name:} mus****hui.com.cn; 
\textbf{IP:} 114.*.*.114; 
\textbf{Port:} 53; 
\textbf{Domain:} LJHD; \ldots \\ 
Relationship & 
\textbf{Alert\_description:} Malicious DNS domain resolution; 
\textbf{Tags:} attack.ta0011; 
\textbf{Level:} High; 
\textbf{Time:} 2026-01-02 \\
\bottomrule
\end{tabularx}
\end{table}

As illustrated in Figure~\ref{fig:event_source} (a), the SOC is overwhelmed with thousands of raw alerts daily. Even after applying initial filtering and deduplication mechanisms, the volume of high-priority alerts necessitating further manual investigation remains in the hundreds per day. Our empirical data reveals that 80\% of these actionable alerts originate from cloud infrastructures, 12\% from enterprise Windows workspaces, and the remaining 8\% from managed client security services. Furthermore, as shown in Figure~\ref{fig:event_source} (b), the taxonomy of alert types varies distinctly across these environments. Malicious file execution constitutes the dominant threat within Windows workspaces, whereas cryptojacking (cryptocurrency mining) is the most prevalent attack vector in cloud environments. In contrast, Remote Access Trojan (RAT) infections are the most frequently observed incidents across the remaining managed systems.

To systematically characterize the complexity of attack investigations, we randomly selected 596 alerts spanning all platforms and alert categories. We analyzed these cases across four critical dimensions: attack chain length, temporal span, number of involved attack stages (tactics), and alert recurrence frequency.
Figure~\ref{fig:Overview_of_investigation} (a) illustrates the distribution of attack chain lengths. The data reveals that the majority of attack chains consist of 3 to 10 hops (e.g., the 7-hop scenario presented in Figure~\ref{fig:example} (a)); however, a small subset of highly complex chains exceeds 10 hops. Regarding temporal span, Figure~\ref{fig:Overview_of_investigation} (b) demonstrates that while most attacks conclude within four hours, several outliers persist for over a day, with extreme cases spanning up to a month.
Furthermore, Figure~\ref{fig:Overview_of_investigation} (c) quantifies the number of attack stages (tactics) per chain. Although most chains encompass only 2 to 4 tactics, some sophisticated attacks traverse up to 7 distinct tactics. Finally, Figure~\ref{fig:Overview_of_investigation} (d) analyzes the daily recurrence frequency of these alerts. We observe that over half of the alerts are first-time occurrences, lacking any historical precedent to assist triage. Conversely, a distinct portion of alerts recur for more than seven days. This sustained recurrence strongly suggests that the initial attack entry may be misidentified or improperly contained, allowing the underlying threat to continuously trigger alerts without effective remediation.

\begin{figure}[t]
  \centering
  \includegraphics[width=0.96\linewidth]{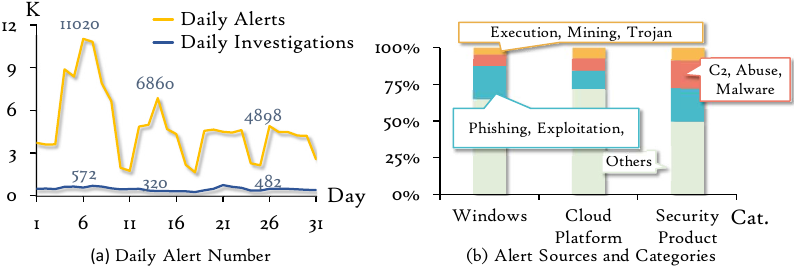}
  \caption{Statistics of Initial Alerts (December, 2025)}
  \label{fig:event_source}
\end{figure}

\begin{figure}[tbp]
  \centering
  \includegraphics[width=0.96\linewidth]{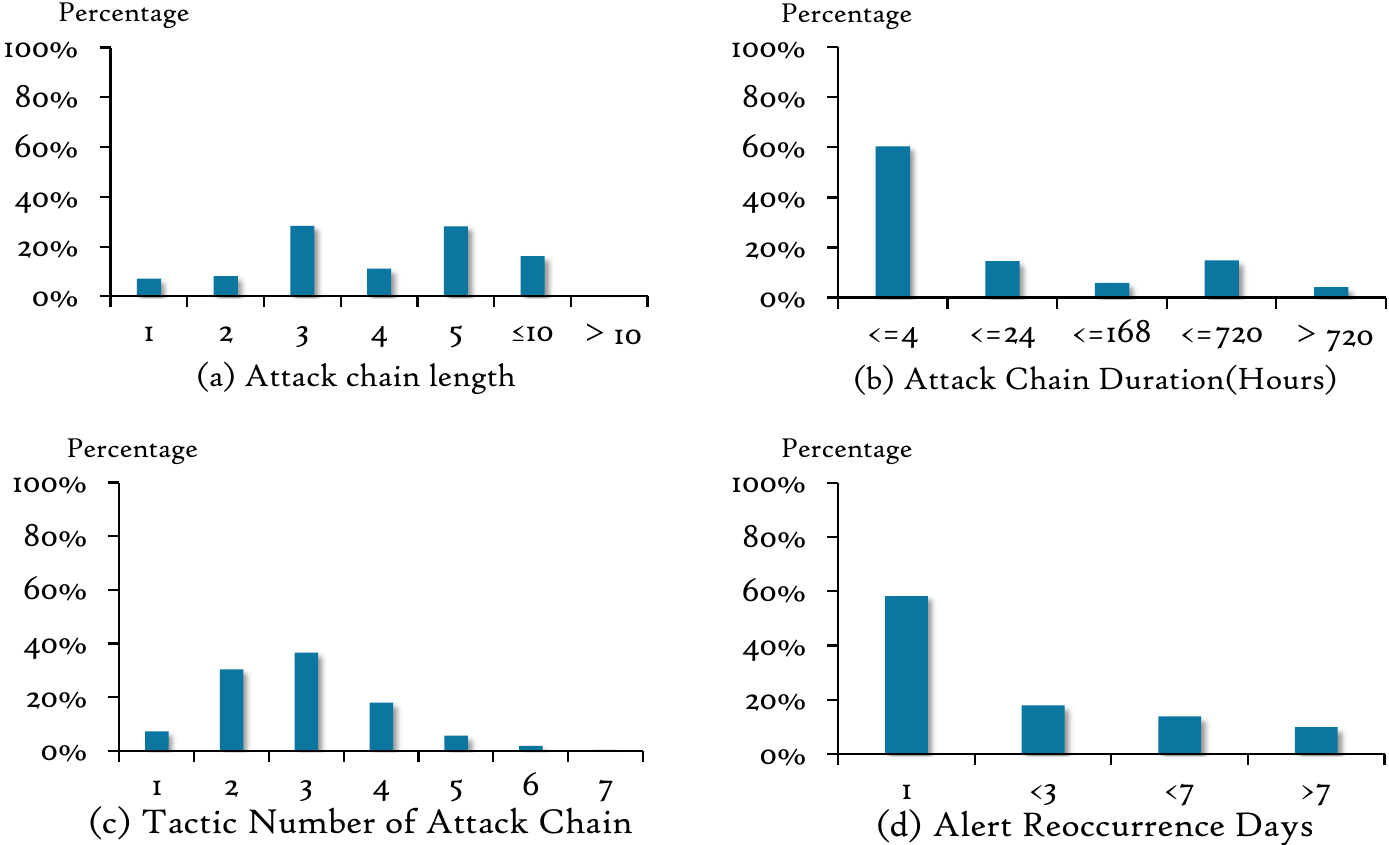}
  \caption{Characteristics of Alerts (Manually Analyzed)}
  \label{fig:Overview_of_investigation}
\end{figure}

In summary, modern SOCs are overwhelmed daily with a massive volume of heterogeneous alerts across multiple sources. The investigation of these alerts is notoriously complex, as they typically feature lengthy causal attack chains, extended temporal spans, multi-stage tactical progressions, and a lack of historical precedent. Furthermore, our empirical user study (\S\ref{eval:userStudy}) indicates that human analysts require between 10 and 20 minutes to triage and investigate a single high-severity alert. This severe operational bottleneck underscores an urgent and critical need for scalable, reliable, and automated attack investigation systems.

\subsection{Motivating Example}
\label{sec:Motivation}

As illustrated in Figure~\ref{fig:example}, an alert is triggered when \texttt{Python.exe} accesses a malicious IP. The investigation reveals that \texttt{aria2.exe} first downloads \texttt{Revit.zip}, which is then decompressed into a markdown (\texttt{.md}) file. This file is subsequently renamed to a Python script and executed, thereby triggering the attack behavior (the POI we captured). This case exhibits fragmented causal chains caused by missing entities and relationships. Specifically, the absence of an execution record for the script file by the \texttt{Python.exe} process prevents traditional backtracking-based investigation systems from reconstructing the complete attack chain. As a result, these systems can only trace back to benign processes, such as \texttt{Explorer.exe}, ultimately leading to investigation failure. Worse still, even without the impact of these broken chains, the lack of semantic analysis capabilities makes it impossible for the investigation to determine which of the multiple entry IPs is the true source of the attack.

The SOC-baseline, which relies directly on an LLM without structured guidance, partially alleviates this issue during the initial alert analysis by identifying script behaviors from the alerts. When no useful log is available from the source process of \texttt{Python.exe}, the baseline attempts to query the script source. However, its subsequent queries retrieve all log fields indiscriminately, preventing the system from consistently focusing on critical fields such as \texttt{CmdLine}. Consequently, even after identifying the script file, the system fails to recognize the renaming action reflected in the \texttt{CmdLine} and cannot trace back to the original \texttt{.md} file. The investigation then degenerates into unguided exploration, which introduces substantial noise, wastes computational resources, and ultimately fails.

Therefore, an ideal system design must carefully define the query scope, avoiding both the additional overhead introduced by an excessively broad scope and the data omission caused by an overly narrow one. Thus, \SysName\ employs provenance-guided causal analysis capabilities, utilizing query templates to restrict the query scope. Simultaneously, it leverages the LLM to flexibly incorporate unstructured information from the \texttt{CmdLine} into the analysis. This approach allows the system to identify script execution behavior and accurately locate the corresponding logs. Through \texttt{CmdLine} analysis, \SysName\ correctly captures the script renaming action. Consequently, \SysName\ infers that the script originated from a renamed markdown file and traces this file back to its creation event, successfully overcoming the broken-chain issue. Furthermore, when identifying the source of the download, \SysName\ utilizes the LLM to analyze full URL semantics rather than relying solely on IP-level information. By analyzing the URL metadata, \SysName\ correctly identifies the true source IP. Detailed adjustments to the query strategy are further discussed in Section~\ref{eval:CaseStudy}.

\subsection{Challenges Faced by the Existing Investigation Systems}
\label{sec:challenge}

\begin{figure}[tbp]
    \centering
    \includegraphics[width = 0.99\linewidth]{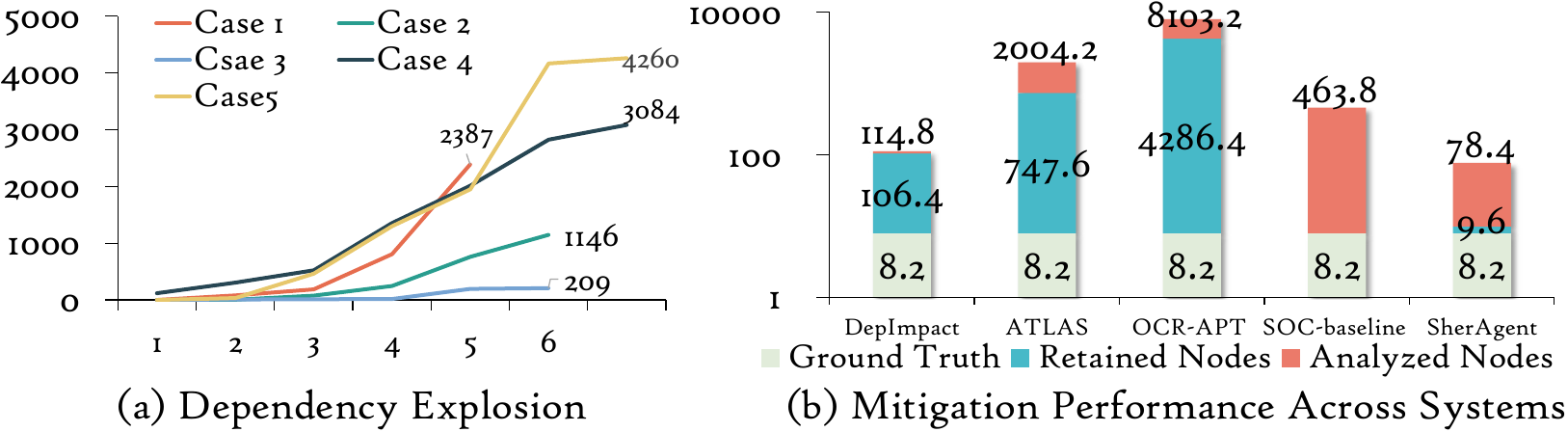}
    \caption{Impact of Dependency Explosion}
    \label{fig:explosion}
\end{figure}

\begin{figure}[tbp]
  \centering
  \includegraphics[width=0.99\linewidth]{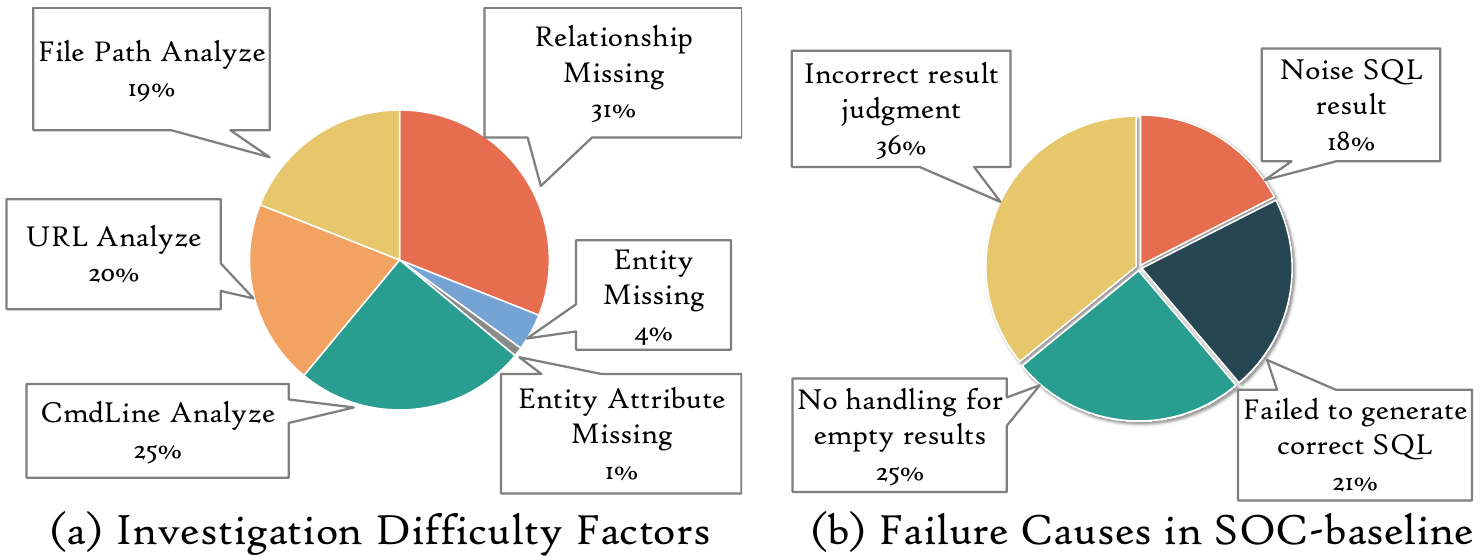}
  \caption{Insufficient Information \& Failure Reason}
  \label{fig:type_fail_combined}
\end{figure}

\subsubsection{Impact of Dependency Explosion}
We randomly select five cases from our experimental dataset that can be successfully resolved by all evaluated systems to ensure comparability. As shown in Figure~\ref{fig:explosion} (a), a dependency explosion occurs when the influence of a single event propagates across an excessive number of system entities~\cite{11169421}, substantially increasing the analyst's burden.
Figure~\ref{fig:explosion} (b) summarizes the total number of queried nodes versus the number of nodes retained in the final investigation results across these five cases. The results indicate that even when a complete attack chain is successfully reconstructed, the volume of nodes requiring analysis varies significantly across different systems. Among them, propagation-based systems like \textsc{DepImpact}~\cite{277080} perform relatively well by computing accumulated weights to filter out irrelevant nodes. Conversely, anomaly-based systems such as ATLAS~\cite{263852} and OCR-APT~\cite{10.1145/3719027.3765219}, which extract anomalous nodes to reconstruct attack chains, inherently require the analysis of considerably more nodes. This is primarily because system behaviors in real-world environments are highly complex, making genuine anomalies much harder to accurately identify.

We also evaluate the SOC-baseline. Since this system typically outputs only the final investigation results, we rely solely on the total number of nodes analyzed during its query phase. As depicted in Figure~\ref{fig:explosion} (b), these existing methods retain a substantial number of irrelevant nodes, indicating that the dependency explosion problem persists. Compared to these systems, \SysName\ significantly reduces the volume of irrelevant logs returned during queries.

\subsubsection{Insufficient Information for Provenance-based Investigation}

The traditional methods struggle with two main aspects: missing log data, which hinders complete investigations due to their strict reliance on causal connectivity, and insufficient semantic analysis, which leads to a high number of false positives.
As shown in Table~\ref{tab:Problem_type}, missing log data can be categorized into missing relationships, missing entities, and missing entity attributes. The primary causes of this information loss include the improper filtering or merging of excessive logs generated by file read/write operations, the inherent difficulty of collecting logs from virtual or external devices, the use of anti-forensic techniques by attackers to erase evidence, or simply incomplete data collection that causes connection building to fail. Ultimately, the loss of this information leads to fragmented attack chains, halting further progress and causing traditional investigations to terminate prematurely.

To overcome these challenges, security analysts often compensate manually. They inspect fields such as \texttt{CmdLine} and entity paths to restore missing relationships, carefully refine SQL predicates and field-matching criteria, infer missing attributes from historical records, and expand queries to parent paths to retrieve related logs and restore correlations. In this work, we attempt to replicate these manual expert strategies using Large Language Models (LLMs) to automatically resolve the aforementioned issues.
Furthermore, insufficient semantic analysis remains a key limitation of traditional methods. As discussed in Section~\ref{sec:Motivation}, this shortcoming primarily leads to false positives due to the inability to accurately distinguish benign activities from genuinely malicious events.

\subsubsection{The limitation of Naive LLM-based Investigation}

Subsequently, we manually analyzed nearly 200 cases where the SOC-baseline (a naive LLM-based investigation approach, with the prompt detailed in Appendix Figure~\ref{fig:prompt3}) failed, summarizing the direct causes of failure and the primary investigative challenges, as illustrated in Figure~\ref{fig:type_fail_combined} (b). First, our analysis reveals that many manually solvable alerts are misanalyzed by the SOC-baseline system due to noisy query results, unstable query generation, the lack of effective adjustment strategies, and incorrect result judgments.

Noisy query results often arise from the inherent permissiveness of SQL matching, which introduces a vast volume of irrelevant logs. This surplus of noise causes the LLM to either overlook critical events or be diverted down incorrect investigation branches. Inappropriate query generation stems from the LLM's lack of domain-specific investigative knowledge. The LLM frequently pursues incorrect investigation directions or applies unreasonable query constraints, leading to ineffective queries and a waste of computational resources. Moreover, the SOC-baseline lacks an effective adjustment strategy for handling uninformative query results. Relying directly on the LLM rarely produces meaningful query refinements, frequently resulting in investigation failures. Finally, this overall lack of investigative knowledge often leads to incorrect final judgments, generating both false negatives and false positives.

We also acknowledge that environmental complexities render certain alerts unrecoverable. Specifically, logs may be discarded due to retention expiration, as extreme data volumes often necessitate the deletion of historical logs. Additionally, URL normalization can hinder investigations, as redirection through unified domains masks specific user activities. For example, the QQ Browser redirects all requests through a unified domain (e.g., \texttt{masterconn.**.com}), obscuring the actual destination URL from the logs.
Overall, these limitations underscore the inadequacy of existing investigation systems, highlighting the need for improved system design.

\begin{figure}[tbp]
    \centering
    \includegraphics[width = 0.97\linewidth]{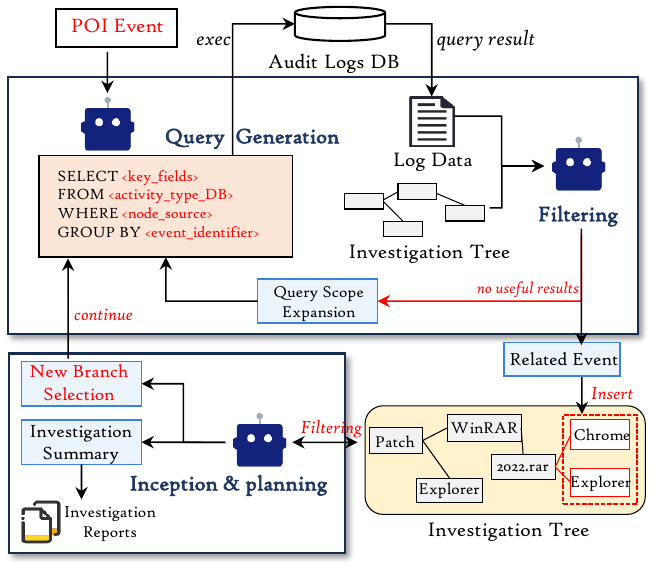}
    \caption{\SysName Architecture}
    \label{fig:architecture}
\end{figure}

\section{\SysName}
\label{sec:design}

\subsection{Overview}

Attack investigation aims to backtrack from a POI and reconstruct the underlying attack chain by identifying relevant entities and their causal relationships. The backtracking process can be conceptualized as an iterative cycle of querying and filtering. However, this task is inherently challenging due to incomplete and noisy log data. During the querying stage, missing or partially observed events often lead to broken chains, hindering reliable connection of related activities. In the subsequent analysis stage, the inclusion of loosely related events can introduce a large number of spurious dependencies, resulting in a dependency explosion that obscures the true attack structure. Therefore, the core challenge lies in striking an optimal balance between causal rigor and query flexibility.



Figure~\ref{fig:architecture} shows the architecture of \SysName. As detailed in Appendix \S\ref{appendix:workflow}, we simulate the security analyst’s workflow by automating the investigation through an iterative query-analysis loop to progressively reconstruct the attack chain. \SysName consists of three core modules: (1) query generation, (2) branch filtering, and (3) result analysis and re-querying.
Specifically, to address the challenge of query generation, particularly in log data omissions, we utilize the LLM to analyze the incoming alert or investigative branch and generate a SQL statement that replaces connectivity-based queries with existence-based queries to mitigate the impact of missing data. To enhance query effectiveness, we provide the LLM with prompts incorporating SQL templates and adjustment strategies. Upon retrieving the queried logs, we perform branch filtering by enriching the log data with semantic fields, thereby enabling the LLM to better interpret the actions of current events. Concurrently, to maintain a long-term state, we store investigative progress in a tree structure. This serves as contextual input for the LLM, facilitating reliable branch filtering. Moreover, to enable iterative investigation, we incorporate an analysis and feedback module. This module continuously updates the tree with filtered events, revalidates causal relationships, and prunes irrelevant branches to ensure the reliability of investigation results. Based on the tree-structured investigative progress, the system determines whether the investigation can be concluded or further exploration is required.

\subsection{Query Generation}
\label{sec:query_generate}


\begin{figure}[tbp]
    \centering
    \includegraphics[width = 0.98\linewidth]{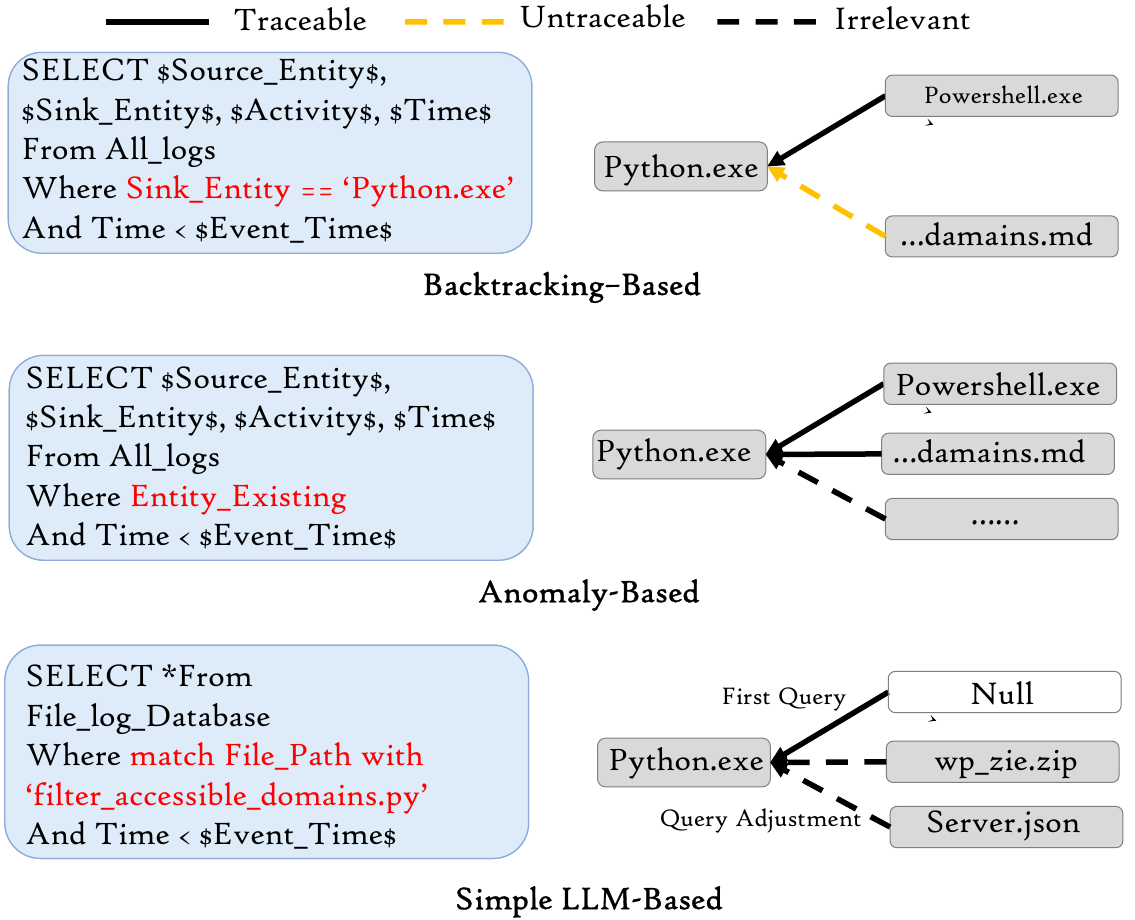}
    \caption{Query Statement and Effect of Existing Methods (Taking the branch in Figure~\ref{fig:example} where the \texttt{Python.exe} process creates the renamed file \texttt{domains.py} as an example)}
    \label{fig:Existing_system}
\end{figure}

As shown in Figure~\ref{fig:Existing_system}, traditional methods are limited by rigid query entity selection and predefined matching conditions. Backtracking-based systems rely on strict connectivity, making them unable to handle broken chains. In contrast, some anomaly-based methods analyze all entities to bypass connectivity constraints, but introduce substantial noise, resulting in high computational overhead and false positives. Furthermore, as illustrated in Figure~\ref{fig:prompt3}, the SOC-baseline lacks an effective strategy, leading to repeated query adjustments and ultimately invalid results.

\begin{figure}[tbp]
    \centering
    \includegraphics[width = 0.94\linewidth]{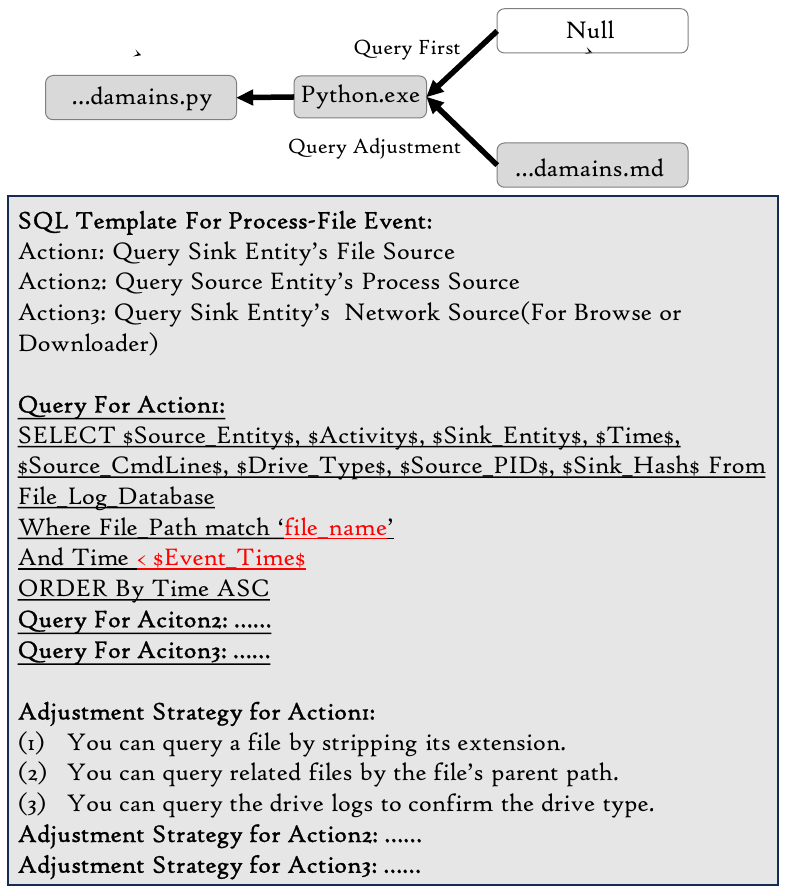}
    \caption{Prompt for Query Generation}
    \label{fig:prompt1}
\end{figure}

Therefore, we design a series of SQL templates based on causal analysis. As illustrated in Figure~\ref{fig:prompt1}, when receiving an alert or branch event, \SysName identifies query entities and activity types. Specifically, we delimit the query scope for distinct event types to refine the search space and ensure targeted data retrieval, while removing the unnecessary fields from query results, like the EDR deployment time and invalid UIDs, to reduce noise. When querying a specific entity, \SysName only needs to fill specific matching conditions, such as entity name and temporal constraints. Notably, \SysName leverages the LLM’s contextual reasoning to discern the investigative action, rather than categorizing queries solely based on event actions. For instance, a read operation during a data exfiltration event necessitates tracing the originating process to identify the malicious actor, whereas the same read operation during file decompression requires tracking the file source to establish the provenance of the archive. These distinct priorities lead to different investigation directions despite identical activities.

Furthermore, we prioritize events that are more likely to be attack-relevant based on LLM reasoning. Specifically, process and network activities are prioritized according to temporal proximity, as events occurring in close succession are more likely to be correlated. Conversely, for file activity, which offers higher query precision through file name, the system sequences the earliest logs first to establish provenance, while subsequent logs are provided as supplementary context for intermediate steps. In scenarios where a query yields no useful results, we implement stepwise adjustment strategies to progressively expand the query scope. We provide guidance-driven adjustment strategies tailored to different source type query actions. As shown in Figure~\ref{fig:prompt1}, for file source query, the query scope can be expanded by stripping extensions or matching the parent path. This allows \SysName to find more related file logs. If these adjustment strategies remain unsuccessful, \SysName abandons the current branch and redirects its focus to the selection of alternative investigative branches.


\subsection{Branch Filtering}


{Dependency explosion remains a core challenge in attack investigation.} Traditional heuristic and anomaly-based methods lack semantic depth, failing to distinguish benign activities from malicious noise and leading to irrelevant branch selection. While LLM-integrated SOC systems introduce semantic parsing, they are hindered by undifferentiated attention across fields due to a lack of structured guidance coupled with the loss of essential context as protracted investigations exceed native model memory.

To address these issues, we incorporate essential semantic fields into the logs, such as CmdLine, file path, etc., while explicitly defining critical fields within the prompt instructions, enabling \SysName to better understand the events under analysis.
To sustain the long-term efficacy and mitigate memory loss for the LLM, we maintain an investigation tree that records analytical progress and serves as the primary provider of contextual information for branch filtering. Each edge in this tree is a tuple of <\textit{source entity}, \textit{sink entity}, \textit{relationship}, \textit{timestamp}>, further enriched with supplementary metadata including full execution paths, CmdLine, drive\_type, and dynamic state tags, specifically ``Used'' for completed branches, ``Tracing'' for the branch currently under analysis, and ``New'' for unexplored branches. This structured management allows the system to maintain a persistent investigative context. For instance, if the tree captures that \texttt{Apex.exe} was spawned by \texttt{Msiexec.exe}, the system can intelligently prioritize subsequent queries toward MSI installation packages.
Finally, \SysName selects up to top-$K$ related events that can further advance the investigation. As evaluated in Appendix \S\ref{appendix:k}, we set $K=3$ based on empirical results.

\subsection{Results Analyzing and Re-Querying}
\label{sec:Dead-Handing}
After branch filtering, we will update the related events to the investigation tree and mark them as new branches. To ensure structural coherence, the system enforces a connectivity constraint where all newly added branches must link to the current \texttt{Tracing} branch. In scenarios involving broken chains where direct connectivity is missing, \SysName proactively bridges the gap by establishing a specialized edge with an action labeled as ``potential correlation'' to connect the new branch with the \texttt{Tracing} path. 
Then, as shown in Figure~\ref{fig:prompt2}, the system performs a global audit of the entire tree to verify whether newly added edges reflect valid causal relationships, pruning those that are irrelevant. This global verification allows \SysName to determine whether the current evidence is sufficient for a final report or whether further queries are needed, effectively replicating a human analyst’s reasoning.

\begin{figure}[tbp]
    \centering
    \includegraphics[width = 0.94\linewidth]{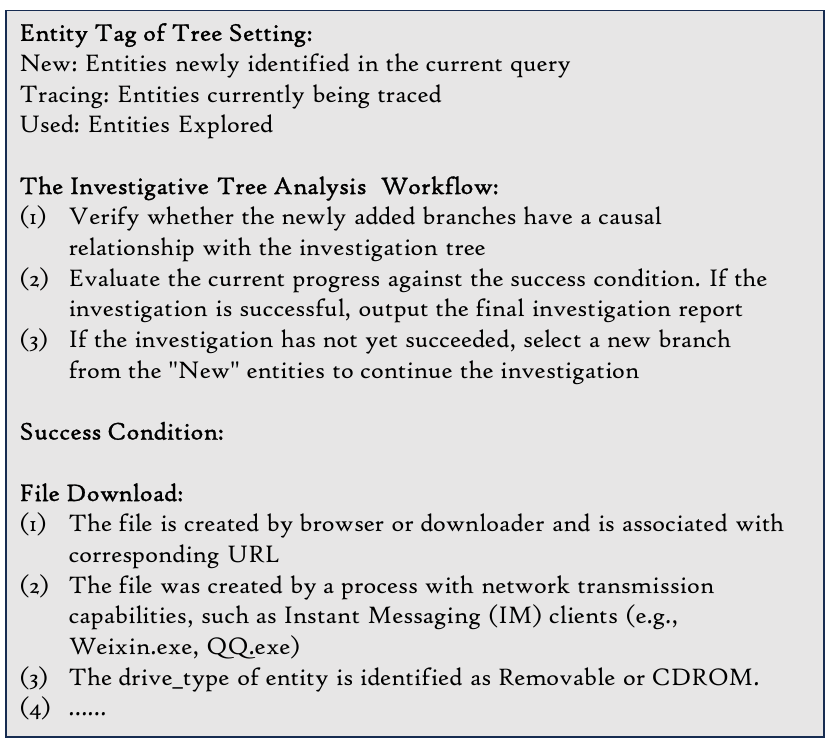}
    \caption{Prompt of Investigative Tree Analysis}
    \label{fig:prompt2}
\end{figure}
Traditional methods lack explicit termination criteria and merely output ambiguous intermediate states when stalled. Furthermore, direct reliance on the LLM for judgment lacks stability as the same case may yield inconsistent or even erroneous results. To ensure reliable termination judgment, we synthesized all attack investigation reports since December 2025 into five primary categories: file download, exploitation, brute force, remote execution via security agents, and supply chain poisoning. We also established specific determination conditions for these types. For example, as shown in Figure~\ref{fig:prompt2}, one indicator of file download is that the file is created by a process with network transmission capabilities. By leveraging these initial access descriptions for comparative analysis, \SysName can reasonably determine the investigation outcome. A successful match signifies a completed investigation. 
If the attack entry is not yet localized while unexplored branches remain in the investigation tree, the system selects the most critical branch for the next investigation step. Conversely, if no match is achieved and all investigative branches have been exhausted, the system records an investigation failure.
This entire investigation culminates in the generation of a comprehensive investigation report and the final state of the investigation tree as a clear record that enables security analysts to quickly understand and verify the investigation result.

\section{Evaluation}
\label{sec:evaluation}

In this section, we focus on evaluating the performance of \SysName by answering the following research questions (RQs).


\begin{enumerate}[label=\textbf{RQ\arabic*.}, left=0pt, itemsep=0em]
\item How effective is \SysName in conducting attack investigations? (\S\ref{eval:effectiveness})
\item How robust is \SysName in mitigating broken attack chains and dependency explosions? (\S\ref{eval:effectiveness}) 
\item How does the foundational LLM impact the overall efficacy and accuracy of \SysName? (\S\ref{eval:overhead}) 
\item Can \SysName enhance the efficiency of manual security analysis? (\S\ref{eval:userStudy})
\end{enumerate}

\subsection{Evaluation Setup}

\begin{table*}[t]
\centering
\caption{Comparison with SOTA Investigation Systems (SR: Success Rate, IR: Incorrect Rate. Triple-values (e.g., $x/y/z$) represent results across three independent trials.)}
\label{tab:Investigation_Result}
\small
\begin{tabular}{l|cc|cc|c|c}
\toprule
{System} & \makecell{\# of Successes \\ (Total: 125+50)} & \makecell{\# of Incorrect \\ {Results}} & {SR} & {IR} & \makecell{Hallucination Success in \\ Uninvestigable Cases  (Total: 25)} & {Consistency} \\
\midrule
\textsc{DepImpact}~\cite{277080} & 0 + 50 & 166 & 28.5\% & 94.8\% & 8 & / \\
ATLAS~\cite{263852} & 9 + 22 & 175 & 17.7\% & 100.0\% & 25 & / \\
OCR-APT~\cite{10.1145/3719027.3765219} & 26 + 13 & 172 & 22.3\% & 98.2\% & 25 & / \\
SOC-baseline & 65/67/74 + 37/37/41 & 73/71/60 & 61.1\% & 38.9\% & 22/18/17 & 66.5\% \\
{\SysName} & {113/113/110 + 50/48/50} & {12/14/15} & {92.2\%} & {7.8\%} & {3/2/2} & {88.5\%} \\
\bottomrule
\end{tabular}
\end{table*}

Our experiments were conducted on the internal cloud computing platform. We developed \SysName using Python. For deployment, we used DeepSeek-V3.1 provided by SOC with max completion tokens set to 64000 and temperature set to 0.6, enabling sufficiently long reasoning traces while maintaining a moderate level of stochasticity. We used ClickHouse to store and manage logs, organizing them into three databases for process, file, and network activities.

\noindent\textbf{Dataset.} 
{\SysName has been deployed in the SOC's production environment and has processed 53,849 alerts for attack investigation. While we recorded the general investigation outcomes for all production alerts on the live \SysName system, the immense volume of data and the prohibitive cost of adapting all baseline systems, combined with the labor-intensive nature of manual ground-truth annotation, preclude a full-scale comparative study on the entire dataset. Therefore, we selectively extracted several groups of representative cases for a more granular evaluation.} Specifically, since 11.4\% of cases suffer from log omissions (as discussed in Section~\ref{sec:challenge}), we identified and extracted 125 cases from this subset. Furthermore, some cases are uninvestigable, yet the SOC-baseline reports them as successful. To evaluate the reliability of systems, we specifically selected 25 such cases where the SOC-baseline generated hallucination success. We note that this selection is biased toward adversarial cases for the SOC-baseline. To assess \SysName's reliability more broadly, we further report the statistics of our daily spot-check results in Section ~\ref{eval:userStudy}. Finally, we randomly chose 50 cases with complete log data to provide a comprehensive benchmark for comparing attack investigation capabilities across all systems. {For all selected cases, we manually annotated the ground truth and further validated the annotations with three SOC security analysts. The final labels were determined through majority voting after discussion to ensure consistency and reliability.}

\noindent\textbf{Baseline.}
We evaluate \SysName against three state-of-the-art (SOTA) attack academic investigation systems, as well as an SOC-baseline (LLM-based) investigation method used as a baseline. Specifically, \textsc{DepImpact}~\cite{277080}, ATLAS~\cite{263852}, and OCR-APT~\cite{10.1145/3719027.3765219} represent attack investigation systems based on backtracking-based, heuristic-based detection, and anomaly-based detection, respectively. For systems that require training, we construct dedicated training datasets. Specifically, for ATLAS, we select 10 cases as the training set, including log data omissions, uninvestigable cases, and cases with complete log data. For OCR-APT, we partition logs within each case temporally, using the first appearance of the attack entry point as the boundary to split the data into training and testing sets.

\begin{figure}[tbp]
    \centering
    \includegraphics[width = 0.98\linewidth]{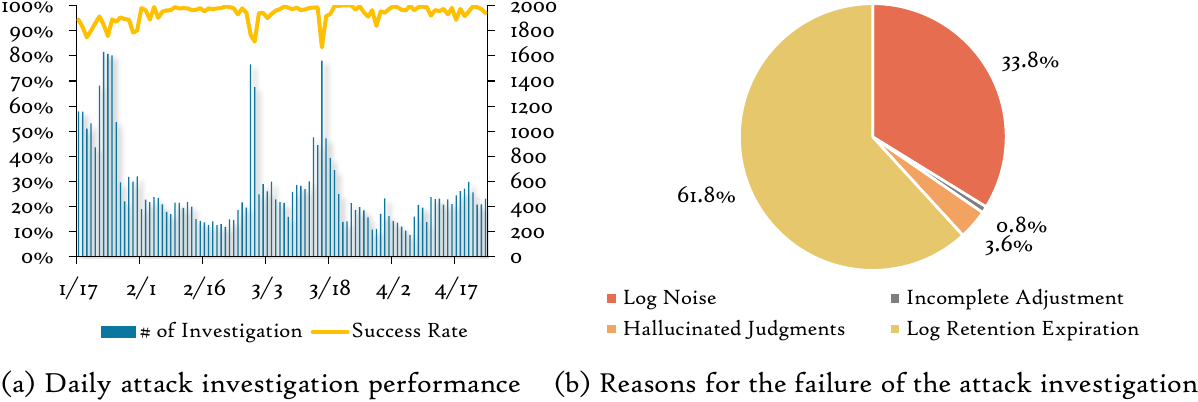}
    \caption{Performance of \SysName in a real-world SOC environment} 
    \label{fig:Application2}
\end{figure}

\begin{figure}[tbp]
    \centering
    \includegraphics[width = 0.91\linewidth]{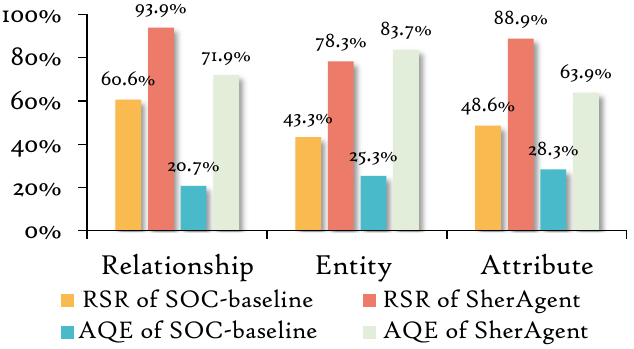}
    \caption{Comparison on Broken Chain Cases (RSR: Recovery Success Rate; AQE: Average Query Effectiveness) }
    \label{fig:broken_chain}
\end{figure}

\subsection{Attack Investigation Effectiveness}
\label{eval:effectiveness}
As illustrated in Figure~\ref{fig:Application2}, \SysName has maintained a consistently high daily investigation success rate. Based on periodic spot-checks—sampling 10 cases per day for a total of 980 cases (approximately 2\% of all alerts)—the human-verified success rate has exceeded 96\% since deployment.
On specific days, high-frequency repetitive attacker activities, particularly in cloud platforms, caused a localized decrease in success rates. Manual analysis revealed that the earliest related events for these redundant alerts had already reached their retention limits, making it impossible to trace the original attack source. Such repetitive alerts not only inflated the daily alert volume but also reduced the overall investigation success rate for those periods.

To better evaluate the attack investigation effectiveness of \SysName, we compared it with \textsc{DepImpact}, ATLAS, OCR-APT, and the SOC-baseline.  {We define the Success Rate (SR) as the ratio of correctly identified entry points to the total number of investigable cases, while the Incorrect Rate (IR) is defined as the proportion of cases where an incorrect entry was selected or the investigation failed to produce a valid result, over the same set of cases. To ensure a conservative and fair comparison, we adopted a lenient success criterion for traditional baseline methods, whereby an investigation is recorded as a success as long as the correct entry is identified, even if the system simultaneously generates incorrect entries. Under this strategy, a single case for a baseline system may contribute to both success and incorrect counts.}

Regarding \textbf{broken chains}, as shown in Table~\ref{tab:Investigation_Result}, \textsc{DepImpact} failed entirely due to its dependency on log continuity, while ATLAS identified 9 cases by relaxing edge directionality. Although OCR-APT employs node-level anomaly detection to bypass broken chains, its training-based method struggles with the complex real-world environments, leading to incomplete detection of attack entities and yielding incorrect results. In contrast, \SysName exhibits high adaptability to diverse broken-chain cases. As shown in Figure~\ref{fig:broken_chain}, {we evaluate this using the ratio of successful retrieval cases 
($ \mathrm{RSR} = \frac{\text{successful retrieval cases}}{{N}} $) 
and average query effectiveness per case 
($ \mathrm{AQE} = \frac{1}{N}\sum_{i=1}^{N} \frac{\text{effective queries}_i}{\text{total queries}_i} $) across three broken chain types.} Here, N denotes the total number of broken chain cases (i.e., N=125). For Relationship Missing, Entity Missing, and Attribute Missing cases, \SysName achieves RSRs of 93.9\%, 78.3\%, and 88.9\%, respectively, significantly outperforming the SOC-baseline. Notably, our system maintains substantially higher AQEs (e.g., 83.7\% in Entity Missing cases compared to the baseline's 25.3\%). This indicates that our SQL templates and adjustment strategies enable more precise query generation, thereby minimizing redundant queries and accelerating attack investigation. In addition, for the LLM, we evaluate consistency, defined as the proportion of cases where three independent runs yield identical results. \SysName achieves a consistency of 88.5\%, significantly outperforming the SOC-baseline's 66.5\%. The occasional inconsistencies in \SysName typically stem from the inherent noise present in certain query outputs. The noise can distract the LLM from the primary attack branch, leading to divergent judgments across different runs. Notably, the SOC-baseline’s consistency is partly inflated by uniform investigation failures on certain cases, leading to apparent consistency.

Furthermore, without semantic analysis capabilities, traditional methods exhibit suboptimal performance even in cases with complete log data. As shown in Table~\ref{tab:Investigation_Result}, these methods frequently fail to distinguish malicious activities from benign ones, leading to a high incorrect rate, and even in cases that are not actually investigable, they still produce spurious attack entries. In contrast, Table~\ref{tab:Investigation_Result} demonstrates that \SysName achieves an SR of 92.2\%, representing an absolute improvement of 31.1\% over the SOC-baseline (61.1\%) and 63.7\% compared to the academic traditional method (28.5\%), while maintaining a low IR of 7.8\%. Moreover, for uninvestigable cases, the system explicitly outputs unsuccessful results rather than fabricating spurious attack entries, further demonstrating the reliability of the results in most cases. These results underscore that \SysName not only ensures high coverage in broken chains but also excels in accurately pinpointing the true attack entries, effectively mitigating the noise. {We performed a root-cause analysis on unique incorrect cases from both experimental and production environments to identify the primary factors behind incorrect results. As shown in Figure~\ref{fig:Application2} (b), the primary cause is log retention expiration (61.8\%), especially in cloud environments where the massive volume of log data necessitates limited retention windows, preventing recovery of long-dwell attacks. This is followed by log noise (33.8\%), which can obscure correct branches due to the matching characteristics of SQL queries. Hallucinated judgments (3.6\%) arise from occasional subjective inferences by the LLM. Notably, even when hallucinated results are correct, we conservatively exclude them from success counts. Finally, incomplete adjustment coverage (0.8\%) reflects the limitations of our static tuning strategy.}

Regarding \textbf{dependency explosion}, we first analyzed their ranking performance for critical investigative nodes on cases they successfully investigated. For traditional methods, effectiveness is measured by ranking ground-truth attack nodes within the Top-$k$ results. In contrast, for \SysName, we measured the ranking of related events during our semantic branch filtering. As shown in Table~\ref{tab:rank}, compared to the best-performing traditional method, \textsc{DepImpact}, \SysName improves the Top-1 hit rate by 81.2\% (from 14.2\% to 95.4\%) and achieves a 2.19$\times$ increase in Top-5 accuracy. This indicates that our semantic enrichment enables the system to reliably identify the optimal investigative path at each decision point. Notably, the SOC-baseline is excluded because it lacks structured branch management and does not generate ranked alternatives for evaluation.
Furthermore, {we quantify dependency explosion mitigation by comparing analytical workload and final retained scale. For a fair comparison, traditional method scales exclude nodes with anomaly scores lower than the ground-truth attack entry. As shown in Table~\ref{tab:rank}, \textsc{DepImpact}, ATLAS, and OCR-APT analyze 199, 9,389, and 10,292 nodes, respectively, while retaining 191, 1918, and 1168 nodes after filtering. For the SOC-baseline, which relies on iterative query execution rather than explicit graph structures, we measure workload by queries per case and nodes analyzed per query. On average, it requires 13 queries per case and processes 46 nodes per query, resulting in substantial redundancy. In contrast, \SysName significantly reduces both the analytical workload and the retained scale, analyzing approximately 136 nodes on average while retaining only 14 nodes in the final investigation tree. This drastic reduction stems from our dual-safeguard mechanism: semantic analysis filters irrelevant events, while investigation tree audits prune redundant branches. These results demonstrate that \SysName effectively neutralizes dependency explosion, while ensuring that critical nodes maintain the highest priority.}

\subsection{Impact of LLM Selection}
\label{eval:overhead}

\begin{figure*}[htbp]
  \centering
  \includegraphics[width=0.89\linewidth]{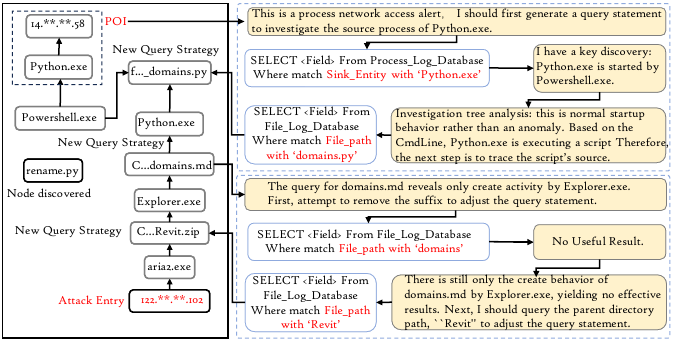}
  \caption{How \SysName Adjusts Query Statements?}
  \label{fig:case_study}
\end{figure*}

\begin{table}[t] 
\centering 
\caption{Filtering Accuracy and Investigation Scale (n/q: nodes/query)} 
\small 
\label{tab:rank} 

\begin{tabular}{l|ccc|c|c} 
\toprule 

System & Top-1 & Top-3 & Top-5 & Analysis & Retained \\ 

\midrule 

\textsc{DepImpact} & 14.2\% & 22.8\% & 45.7\% & 199 nodes & 191 nodes\\ 
ATLAS & 0\% & 22.5\% & 29\% & 9389 nodes & 1918 nodes \\ 
OCR-APT & 2\% & 2\% & 4\% & 10292 nodes & 1168 nodes \\ 

SOC-baseline 
& - & - & - 
& 13 queries &  46 n/q \\ 

\SysName & 95.4\% & 100\% & 100\% & 136 nodes & 14 nodes \\ 

\bottomrule 
\end{tabular} 
\end{table}

In this section, we evaluated the impact of foundational LLMs on \SysName and the SOC-baseline. As shown in Table~\ref{tab:model_comparison}, the Success Rate (SR) of \SysName remains highly consistent across all tested LLMs, with only minor fluctuations.
Even under the weakest-performing LLM within our framework (Kimi-K2.5 at 85.1\%), \SysName still significantly outperforms the strongest SOC-baseline configuration (DeepSeek-V3.1 at 61.1\%), demonstrating a clear and consistent performance advantage across different model scales. These results suggest that the improvements of \SysName are not driven by the capability of any specific model, but by our design in structuring and guiding the investigation process.
Furthermore, we evaluate system efficiency in terms of investigation time, token consumption, and monetary cost. When using the best-performing model, \SysName reduces investigation time by 52.9\%, decreases token usage by 32.3\%, and saves \$0.06 per investigation. Notably, for the self-hosted Qwen3.5-397B-A17B, we exclude monetary cost.

\begin{table}[t]
\centering
\caption{Impact of Foundational LLM Selection on Investigation Effectiveness and Efficiency}
\small
\begin{tabular}{c|c|cc|c}
\toprule
SOC-baseline  & SR & Token & Cost(\$) & Time(s) \\
\midrule
DeepSeek-V3.1      & 61.1\% & 560k & 0.16 & 393 \\
DeepSeek-V3.2  &  55.4\%   & 724k & 0.20 & 400 \\
Kimi-K2.5                   &    59.7\%  & 459k& 0.13 & 148 \\
Qwen3.5-397B-A17B           &   58.2\%    & 438k & / & 157 \\
\midrule
\SysName & SR & Token & Cost & Time \\
\midrule
DeepSeek-V3.1      & 92.2\% & 379k & 0.10 & 185 \\
DeepSeek-V3.2  & 91.4\% & 482k & 0.13 & 358 \\
Kimi-K2.5                   & 85.1\% & 328k & 0.09  & 109 \\
Qwen3.5-397B-A17B           & 85.7\% & 268k & / & 137 \\
\bottomrule
\end{tabular}
\label{tab:model_comparison}
\end{table}

\subsection{Case Study}
\label{eval:CaseStudy}



In Figure~\ref{fig:case_study}, we demonstrate two iterative investigation loops in \SysName, new branch selection and query adjustment strategies.

\textbf{Branch Selection.} Given an alert on the network access behavior of \texttt{Python.exe}, \SysName first queries its process source to identify potential upstream processes. From the retrieved results, the analysis reveals that  \texttt{Python.exe} is spawned by  \texttt{Powershell.exe}, which in turn originates from  \texttt{Windsurf.exe}. These entities are incorporated into the investigation tree for further reasoning. However, although  \texttt{Powershell.exe} is not irrelevant, it corresponds to benign system activity. At this point, instead of continuing along the process lineage, the system selects a new investigation branch. By incorporating CmdLine analysis, the system identifies script execution semantics and infers a hidden branch, namely tracing the origin of the executed script. This leads to querying the script file, ultimately revealing its creation event.

\textbf{Query Adjustment.} When querying markdown file, the system identifies that it was created by  \texttt{Explorer.exe}. We first try to remove the extension to expand query scope. However, this still fails to return relevant results, such as decompression or renaming events. In response, the system relaxes the query constraints by shifting to a higher-level attribute, specifically the parent file path  \texttt{Revit}. This relaxed query successfully uncovers both the archive file  \texttt{Revit.zip} and its creation event by the downloader \texttt{aria2.exe}.

\subsection{User Study}
\label{eval:userStudy}

\begin{table}[tbp]
\footnotesize
    \centering
    \caption{User Study Questionnaire (Partial)} \label{tab:questionnaire}
    \begin{tabular}{p{5cm}|p{2.7cm}}
        \toprule
        \multicolumn{1}{c|}{Questions} & \multicolumn{1}{c}{Options}   \\ 
        \midrule
        How many years of professional experience? &  Less than 1-3 / 4-6 / 7+\\
        How much time is needed to investigate an Alert? & 1-10 / 10-30 / 30+ (min) \\
        What factors contribute to the time consumption during attack investigation? & Broken attack chain /\newline Writing query statement/ \newline Excessive logs / Others \\
        Does \SysName help reduce investigation time? & Yes/No \\
        Approximately how much time is saved by using the system? & 1-10 / 10-20 / 20+ (min) \\
        Please rate the reports generated by \SysName on ``Accuracy,'' and ``Clarity.'' & Very Low / Low / Medium /\newline High / Very high \\
        Daily sampling inspection feedback & Result accuracy \\
        \bottomrule
        \end{tabular}
\end{table}

\begin{figure}[tbp]
  \centering
  \includegraphics[width=0.99\linewidth]{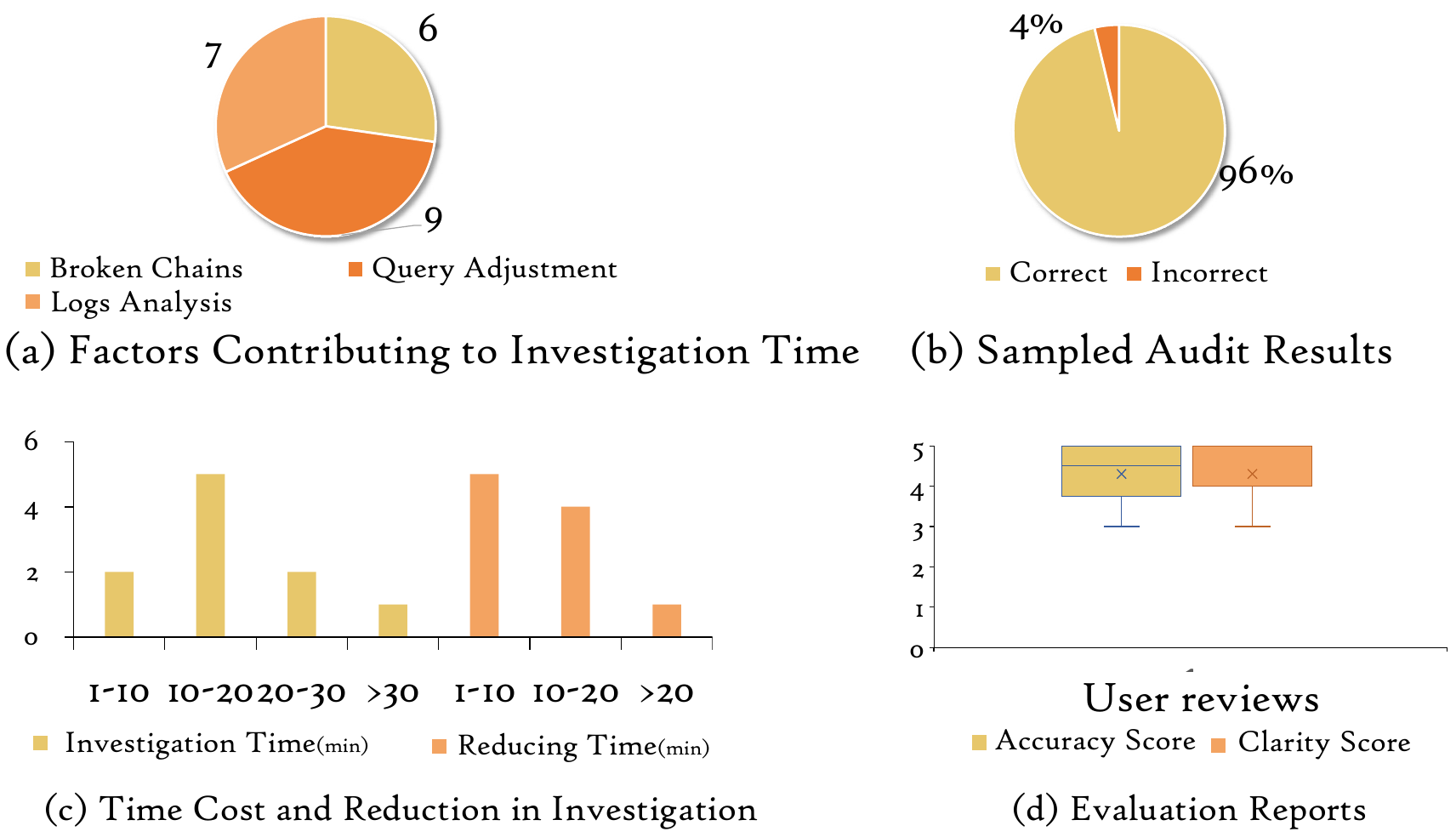}
  \caption{Summary of User Study}
  \label{fig:user_study}
\end{figure}

We conducted a user study with 10 experienced security practitioners from the SOC to evaluate \SysName’s real-world impact. {These participants were selected because they perform daily spot-checks of investigation results generated by our system as part of their daily SOC tasks, ensuring their familiarity with \SysName outputs and making them well-suited for evaluating system performance.} As shown in Table~\ref{tab:questionnaire}, we characterize their investigative experience and evaluate the efficiency and quality of investigations with \SysName. A key outcome of the study was identifying the primary bottlenecks that impede manual investigation. Participants consistently highlighted three major challenges: broken chains, query adjustment, and log analysis. 

\SysName is designed to directly address these bottlenecks. The LLM-based SQL query generation mitigates the impact of broken query chains, while semantic-aware log filtering improves both the efficiency and accuracy of analysis. In addition, our adjustment strategies and branch backtracking mechanism enable \SysName to autonomously rectify invalid queries. Regarding operational efficiency, 50\% of participants reported that \SysName reduced their investigation time by more than 10 minutes (Figure~\ref{fig:user_study} (c)). This highlights a substantial improvement in operational efficiency over conventional practices. 
Moreover, based on participant feedback on accuracy and clarity (Figure~\ref{fig:user_study} (d)), \SysName receives a high average rating of $4.3/5.0$, indicating that participants perceive the generated investigations as both accurate and clear. 

\section{Related Work}

\subsection{Provenance-based Attack Investigation}
Provenance-based investigation analyzes attacks by modeling system audit logs as provenance graphs, where nodes represent system entities and edges represent their causal relationships.
By tracing causal dependencies between entities, such as files, processes, and network events, analysts can reconstruct attack chains from these logs. Existing methods fall into two categories: backward-tracking-based and query-based methods. Specifically, the backward-tracking-based method can be further divided into propagation-based methods and anomaly-based methods.

Propagation-based causal analysis is an early line of work. King et al. proposed BackTracker~\cite{10.1145/945445.945467}, which traces event chains backward from alerts to construct dependency graphs, later extended to cross-host analysis~\cite{king2005enriching}. However, coarse-grained analysis leads to dependency explosion. To mitigate this, subsequent methods adopt fine-grained causal analysis~\cite{Kwon2018MCIM, Lee2013HighAA}, incorporating predefined rules and policies to filter irrelevant dependencies~\cite{10.1145/3133956.3134045,Liu2018TowardsAT}. While improving efficiency, these methods suffer from limited generalization and may discard useful context. \textsc{DepImpact}~\cite{277080} further introduces discriminative dependency weighting to identify attack-related events. However, rule- and feature-driven approaches lack semantic understanding, making it difficult to capture event intent or distinguish adversarial behaviors that mimic benign patterns. Moreover, their reliance on graph connectivity causes the investigation to stall when entities or relationships are missing.

Anomaly-based methods identify attack chains by detecting anomalous events and correlating them to infer attack behaviors. Representative systems, including DEEPCASE~\cite{9833671}, NoDoze~\cite{Hassan2019NoDozeCT}, DeepLog~\cite{10.1145/3133956.3134015}, and RAPIDLE~\cite{wei2026rapidle}, perform event-level anomaly scoring to infer attack paths. ATLAS~\cite{263852}, AIRTAG~\cite{291066}, and OCR-APT~\cite{10.1145/3719027.3765219} further reconstruct attack chains by detecting anomalous nodes using training-based methods. However, these approaches depend heavily on detection accuracy and training data coverage, limiting their effectiveness in real-world environments.

Regarding the query-based methods, Gao et al.~\cite{217496,215975} proposed using domain-specific query languages to extract relevant events from system monitoring or audit logs for attack reconstruction, while concurrently enhancing query efficiency. However, these approaches rely on manually crafted query statements, limiting their practicality. Subsequently, ThreatRaptor~\cite{9458828} leverages open-source cyber threat intelligence to automatically generate queries using external knowledge. However, methods relying on external cyber threat intelligence often struggle to generalize to novel attacks.

\subsection{LLM and Agentic Workflows}

With the rapid advancement of LLMs, their applications in cybersecurity have expanded. In threat detection, researchers integrate LLMs into traditional pipelines to improve detection effectiveness and handle large-scale data processing more efficiently \cite{11113355,10.1145/3658644.3691374,299740}. For vulnerability analysis, LLMs automate discovery across real-world software projects \cite{Lin2025FromLT, inproceedings}. In security report analysis, LLM-based systems extract structured knowledge graphs from CTI reports to create queryable threat representations \cite{10628558,yangCTIThinkerLLMdrivenSystem2026,ZHANG2025104220}. Furthermore, LLMs enable autonomous penetration testing \cite{299699} and the reconstruction of attack behaviors from anomalous subgraphs \cite{10.1145/3719027.3765219}.
Beyond serving as auxiliary tools, the rise of agentic workflows \cite{Yao2022ReActSR, schick2023toolformerlanguagemodelsteach} has empowered LLMs with multi-step reasoning and tool invocation capabilities. \SysName adopts this agentic design to automate query generation and log analysis. By leveraging semantic analysis, \SysName effectively correlates fragmented events and adapts to novel attack scenarios with minimal prior knowledge.



\section{Discussion and Conclusion}
\subsection{Discussion}
\textbf{Environmental Disparities in Real-World Scenarios.} Our investigation of real-world alerts reveals significant gaps between experimental settings and live operations. Production environments frequently suffer from log omissions that break causal chains. Unlike curated datasets, real-world traffic is characterized by massive noise and novel attack patterns that lack the stable training data required by traditional methods, complicating investigation.

\textbf{Limitations.} \SysName struggles to interpret generic system processes, where common CmdLines lack the useful semantic information needed for branch filtering. As discussed in Appendix \S\ref{appendix:three_cases}, in such cases, CmdLine is insufficient, necessitating additional data dimensions. Furthermore, while \SysName handles empty query results, it lacks a robust mechanism to self-adjust when facing excessive noise, such as redundant system logs or self-starting processes. This requires a more autonomous refinement strategy to filter irrelevant data based on query feedback.

\textbf{Advantages in Repeated Attacks and Adaptability.} As shown in Appendix \S\ref{appendix:three_cases}, in repeated attack scenarios, an attacker may replicate entire workflows, including transferring the same malicious file again without reusing the original entry point. While connectivity-based methods are limited to the specific chain linked to a current alert, \SysName uses SQL matching to simultaneously locate multiple semantically related entities across historical instances. Moreover, unlike traditional systems that require labor-intensive normalization to align heterogeneous logs (host, network, and application), \SysName requires only basic structural mapping for SQL queries. By bypassing the need for unified behavioral preprocessing or complex semantic modeling of fields like action, our system offers superior adaptability to diverse and evolving environments.

\textbf{Adversarial Attacks on LLMs.} 
Despite the effectiveness of \SysName, its reliance on LLM reasoning introduces potential vulnerabilities to adversarial attack. We discuss two representative scenarios. First, an attacker may inject misleading information into logs, such as CmdLine, to interfere with LLM reasoning. To mitigate this, program analysis can be used to strip non-functional segments from CmdLines, while the investigation tree enables contextual cross-verification to detect inconsistencies between log content and the established attack path. Second, attackers may employ obfuscation techniques to conceal malicious intent, which poses additional challenges for accurate interpretation. This can be mitigated by integrating mature deobfuscation solutions, such as the subtree-based AST recovery method ~\cite{10.1145/3319535.3363187}, which effectively restores semantics for more accurate analysis.

\subsection{Conclusion}
This paper presents \SysName, an LLM-empowered automated investigation system that employs an iterative "query-filter" backtracking paradigm. By leveraging semantic reasoning, \SysName effectively bridges fragmented causal chains and mitigates dependency explosions. In a real-world production deployment, the system achieved a verified success rate of 96\%, representing an end-to-end improvement of 34.9\% over the enterprise \textit{SOC-baseline} and 67.5\% over academic SOTA methods. Ultimately, \SysName streamlines complex attack investigations into concise, actionable provenance graphs.

\bibliographystyle{ACM-Reference-Format}
\bibliography{refs}

\appendix
\section{Open Science} 

In accordance with the ACM CCS Open Science policy, we provide the following information regarding our research artifacts.

\textbf{Artifact Enumeration:}
To facilitate the evaluation of our paper's core contributions, we provide a \textbf{Functional Evaluation Package}, which includes:
\begin{itemize}[left=0pt, itemsep=0em]
    \item \textbf{Simplified Prototype Code:} A standalone implementation that includes all core functional modules of our system. 
    \item \textbf{Synthetic Datasets:} A collection of synthetic data samples generated to mirror the cases of our real-world datasets, including all broken chain scenarios.
    \item \textbf{Execution Scripts:} We provide reference instructions for environment configuration, along with scripts for data generation and for running the system on synthetic data.
    \item \textbf{Documentation:} A README file providing comprehensive instructions for reproducing our method.
\end{itemize}

\textbf{Reviewer Access:}
The artifacts are available for double-blind review at: \url{https://anonymous.4open.science/r/SherAgent-4A4C}

\textbf{Justification for Non-Shared Artifacts:}
Due to strict corporate confidentiality requirements and the sensitive nature of customer data, certain artifacts cannot be shared in their entirety:
\begin{itemize}[left=0pt, itemsep=0em]
    \item \textbf{Full Production System:} The complete codebase is integrated into our industrial partner's production environment and contains proprietary intellectual property and infrastructure-specific logic that cannot be disclosed.
    \item \textbf{Real-World Datasets:} The raw datasets contain sensitive information from real-world customers. Sharing this data poses significant privacy and legal risks under our data protection agreements.
\end{itemize}

To mitigate these constraints without hindering peer review, we provide a fully functional, simplified version of the code and high-fidelity synthetic data. These artifacts are sufficient for reviewers to verify the algorithmic integrity, performance characteristics, and correctness of our proposed methodology.

\section{Ethical Considerations} 
\textbf{Stakeholders and benefits:} 
This work involves security analysts, the collaborating SOC, the operating organization, end users whose activities are logged, and the broader cybersecurity community. \SysName improves investigation efficiency and accuracy, reducing analyst workload and enabling faster threat response. Organizations benefit from improved incident handling, while the research community gains insights into scalable investigation methods.

\textbf{Ethical principles:}
Following the Menlo Report, beneficence is reflected in improved detection and reduced operational burden. Respect for persons requires careful handling of potentially sensitive log data. Justice concerns arise because benefits may be concentrated in well-resourced organizations. Respect for law and public interest is maintained through compliance with existing SOC data governance practices.

\textbf{Data use and privacy:}
The system operates on enterprise audit logs containing process behavior, file paths, and network metadata. Although such data is standard in security operations, it may include sensitive information. We worked with the collaborating SOC to ensure that all data remained within controlled infrastructure, and identifiable fields (e.g., IP addresses and user-related attributes) were anonymized or minimized where possible.

\textbf{Risks and mitigations:}
Key risks include incorrect investigation results, potential over-reliance on automated outputs, and LLM-related hallucinations observed in prior systems. These are mitigated through structured query constraints, investigation-tree verification, and human analyst oversight via validation and spot-checking. Access to data is restricted within enterprise controls, and sensitive operational details are not disclosed.

\textbf{Decision rationales:}
We proceeded because the security benefits outweigh the risks, and the system operates within established organizational and legal frameworks. While residual risks remain, particularly regarding privacy and dual-use concerns, responsible deployment and disclosure make the work ethically justified.


\begin{figure}[tbp]
    \centering
    \includegraphics[width = 0.98\linewidth]{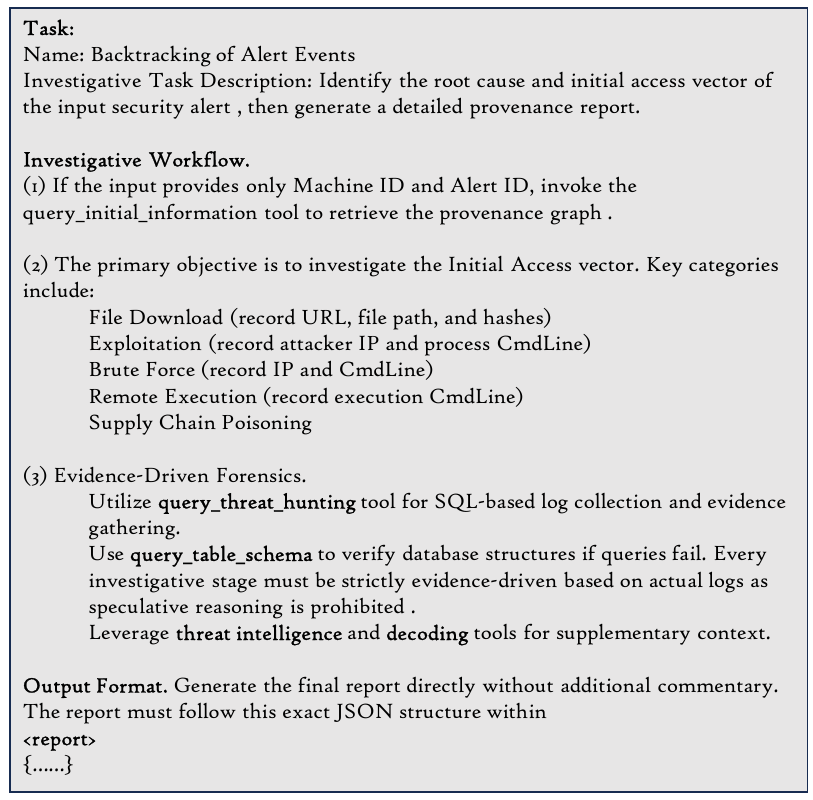}
    \caption{Prompt of the SOC-baseline}
    \label{fig:prompt3}
\end{figure}

\section{Classification and Resolution Strategies for Missing Logs}
As shown in Table~\ref{tab:Problem_type}, log data omissions can be categorized into relationship missing, entity missing, and entity attribute missing.

Relationship missing occurs in three scenarios: excessive file read/write logs that are improperly filtered or merged, uncollected logs from virtual or external devices, and evidence removal via anti-forensic techniques. These issues break chain connectivity and hinder connectivity-based methods from tracing attack entry points. Analysts often compensate by inspecting fields such as CmdLine and entity paths to recover missing relationships.

Entity missing occurs when entities lack associated events, or when event collection failures leave only PID and hash values. However, traditional methods fail to detect such cases. Security analysts manually refine SQL predicates and field-matching criteria to overcome these investigative hurdles.

SOC logs provide auxiliary attributes (e.g., \texttt{drive\_type} for USB devices), which may become unavailable due to device removal or damage. In addition, limited user-mode collection may fail to capture changes in process execution subjects, leading to broken chains. Analysts address these issues by inferring missing attributes from historical records and expanding queries to parent paths to restore correlations.

\begin{table*}[h]
\centering
\footnotesize
\caption{Classification and Resolution Strategies for Missing Logs}
\label{tab:Problem_type}
\begin{tabular}{c|p{4.6cm}|p{4.6cm}|p{4.6cm}}
\toprule
& Relationship Missing
& Entity Missing
& Entity Attribute Missing \\
\midrule
\makecell{Possible\\Reason}
& \makecell[l]{%
(1) Improper filtering or merging of logs\\
(2) Virtual or external devices logs missing\\
(3) Anti-Forensics methods by attackers
}
& \makecell[l]{%
(1) Entity with no associated events\\ (2) Failure collecting event alerting
}
& \makecell[l]{%
(1) Drive removal or damage\\
(2) Limited user mode collection granularity
}
\\
\midrule
Cases
& \makebox[4.6cm][c]{%
    \includegraphics[width=4.4cm]{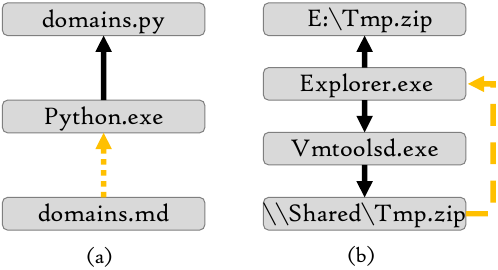}\hfill
}
& \makebox[4.6cm][c]{%
    \includegraphics[width=4.4cm]{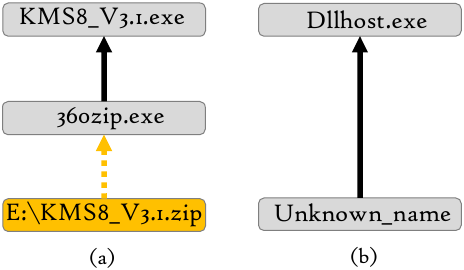}\hfill
}
& \makebox[4.6cm][c]{%
    \includegraphics[width=4.4cm]{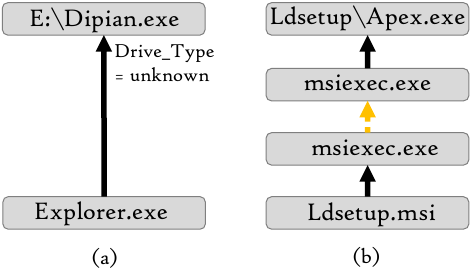}\hfill
}
\\
\midrule
Description
& (a) From CmdLine, we find that \texttt{Python.exe} executed a file renaming script, but the log recording the execution operation is missing.
(b) \texttt{Explorer.exe} copied a file from a shared path to the host, but the log recording the read operation is missing.
& (a) From CmdLine, we find the decompression tool extracted archive \texttt{KMS8\_V3.1.zip}, but all the logs recording the node activity are missing. (b) The source node name fails to be captured in the logs, resulting in garbled identifiers, with only PID and MD5 information retained.
& (a) The file originated from a USB drive, but due to possible user ejection, the associated drive information is missing. (b) The UIDs of the two \texttt{msiexec.exe} processes are different, indicating they belong to separate nodes with no causal relationship.
\\
\midrule
Solution
& (a) CmdLine analysis. Querying md file: \newline
\texttt{SELECT \$Field\$ } 
\texttt{FROM File Database} \newline
\texttt{WHERE \textcolor{red}{File\_path match domain.md}}\newline
(b) No impact. Querying Tmp.zip: \newline
\texttt{SELECT \$Field\$ } 
\texttt{FROM File Database} \newline
\texttt{WHERE \textcolor{red}{File\_path match Tmp.zip}}
& (a) Querying parent path: \newline
\texttt{SELECT \$Field\$} 
\texttt{FROM File Database} \newline
\texttt{WHERE \textcolor{red}{File\_path match E:}} \newline
(b) Querying PID and md5: \newline
\texttt{SELECT \$Field\$ } 
\texttt{FROM Process Database} \newline
\texttt{WHERE \textcolor{red}{Pid == \$Pid\$ and  Hash == \$MD5\$}} 
& (a) Querying parent path: \newline
\texttt{SELECT \$Field\$} 
\texttt{FROM File Database}\newline
\texttt{WHERE \textcolor{red}{File\_path match E:}}\newline
(b) Querying parent path: \newline
\texttt{SELECT \$Field\$}
\texttt{FROM File Database} \newline
\texttt{WHERE \textcolor{red}{File\_path match Ldsetup}} 
\\
\bottomrule
\end{tabular}
\end{table*}

\begin{figure}[tbp]
    \centering
    \includegraphics[width = 0.98\linewidth]{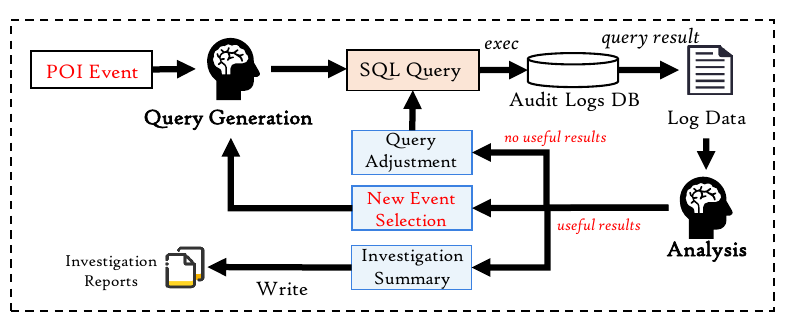}
    \caption{Workflow of the security analyst}
    \label{fig:human}
\end{figure} 

\section{Agents Workflow}
\label{appendix:workflow}
As shown in Figure~\ref{fig:prompt3}, the SOC system provides only a broad investigation workflow along with potential initial access vectors and the required report format. This approach grants the LLM full autonomy over the specific investigative execution. Consequently, the lack of structured guidance leads to the critical shortcomings detailed in Section ~\ref{sec:challenge}. In contrast, \SysName replicates a human analyst's logic through its modular design. As shown in Figure~\ref{fig:human}, when receiving an alert, the security analyst examines the POI and formulates SQL queries to retrieve log data. This investigative phase corresponds to the Query Planner and Generation module. After gathering logs, the analyst filters them to pinpoint relevant evidence, a step executed by the Branch Filtering module. If the investigation stalls, the analyst relaxes query constraints or switches to a new branch. These actions are handled by the Query Planner’s adjustment strategies and the Analysis and Feedback module. Finally, the analyst determines if the attack entry is resolved to generate a formal report. If all potential branches are exhausted without identifying a successful match, the analyst records an investigation failure. These final judgments are performed by the Analysis and Feedback module. Throughout this iterative loop, the Investigation Tree maintains the global state to mimic the analyst's persistent memory.

\section{Selecting the Parameter K}
\label{appendix:k}

\begin{figure}[tbp]
    \centering
    \includegraphics[width = 0.80\linewidth]{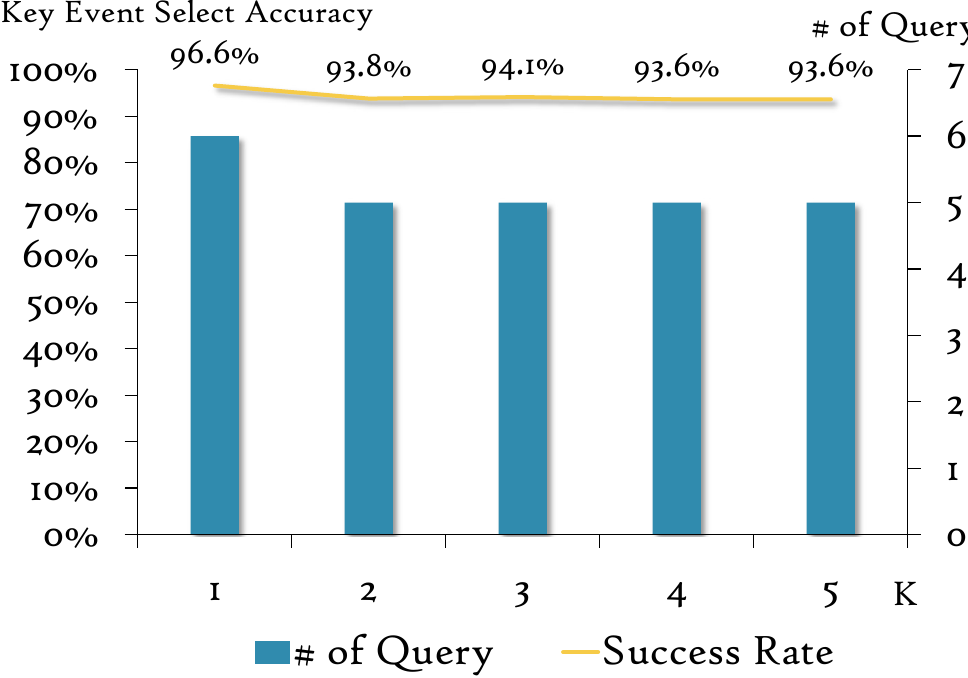}
    \caption{Influence of the Parameter K}
    \label{fig:parameter}
\end{figure}

We evaluated the parameter $K$ that limits the number of filtered log results. As illustrated in Figure~\ref{fig:parameter}, $K=1$ achieves higher accuracy but forces step-by-step investigation, increasing query overhead. Setting $K=3$ strikes a balance by maintaining high filtering accuracy while reducing the number of queries. For instance, a single query can capture multiple operations (e.g., file creation and copying) on the same entity. Larger $K$ values do not further reduce queries but instead harm filtering accuracy. Therefore, we set $K=3$ as a balanced choice.

\begin{figure*}[htbp]
    \centering
    \includegraphics[width = 0.86\linewidth]{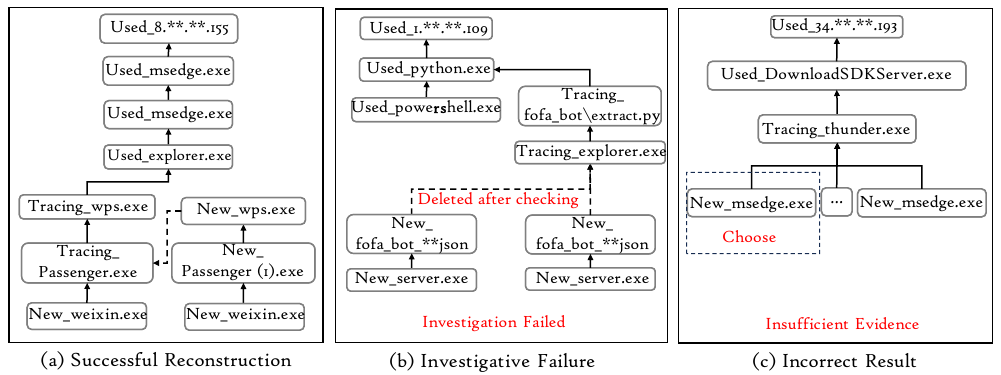}
    \caption{Investigation Trees of Three Cases}
    \label{fig:cases}
\end{figure*} 

\section{Three Cases}
\label{appendix:three_cases}
This section presents three representative case studies to provide a deeper understanding of \SysName in production environments. We visualize these cases as investigation trees to illustrate the system's iterative reasoning process. To clarify the investigative state, each node is labeled with a specific tag. Used represents entities already explored, while Tracing indicates entities currently undergoing active analysis. New identifies entities recently discovered in the latest query iteration. We first examine a successful investigation that effectively manages dependency explosion while successfully capturing recurring attack entries. We then discuss a failure case to analyze the inherent limitations of \SysName. Finally, we analyze a case where the system generates a false positive.

\textbf{Case 1: Capturing Recurring Entries.}
As shown in Figure~\ref{fig:cases} (a), \SysName successfully reconstructed the complete attack chain for this case. During the investigation of the malicious file \texttt{Passenger.exe}, the flexible matching capabilities of our SQL-based query identified additional logs related to a duplicate file copy which is integrated into the investigation tree. Analysis reveals that the attacker or victim performed repeated download and execution activities. The initial occurrence likely remained undetected because the attacker refrained from executing further malicious actions at that stage. Traditional connectivity-based methods fail to correlate such repetitive behaviors. This failure leaves critical security risks unaddressed, as the unresolved root cause often allows for repeated compromises. In contrast, \SysName effectively captures these recurring activities. Although this inclusive recovery results in a node count exceeding the absolute ground truth, these nodes represent valid attack-related associations rather than irrelevant noise.

\textbf{Case 2: Systematic Analysis of Investigative Failure.}
As shown in Figure~\ref{fig:cases} (b), this scenario illustrates a failure in reconstructing the script file source. When direct queries for the script yielded no results, \SysName shifted to the parent path search as part of its adjustment strategy. However, the relevant logs for the \texttt{fofa\_bot} entity are heavily obscured by a massive volume of temporary JSON files. This excessive noise overwhelmed the system and prevented it from successfully locating the original compressed archive. This failure highlights the inherent limitations of relying on static query adjustment strategies. Predefining fixed remediation rules for every possible environmental complexity is impossible. Consequently, future work should focus on developing dynamic adjustment policies that generate tailored query refinements based on the specific context of real-time search results.

\textbf{Case 3: Systematic Analysis of Incorrect Result.}
Figure~\ref{fig:cases} (c) illustrates a case of an incorrect investigation result. After backtracking a malicious network alert from \texttt{DownloadSDKServer.exe} to several \texttt{msedge.exe} processes, \SysName encountered a semantic void. The underlying logs could not definitively map the network call to a specific CmdLine argument among multiple candidates. Faced with this ambiguity, the Analysis and Feedback module adopted a temporal heuristic by selecting the link with the highest time-based proximity. This unverified causal chain results in a qualitative false positive due to insufficient evidence.

\end{document}